\documentclass[12pt,a4paper]{article}
\usepackage{jheppub}
\usepackage[dvipsnames]{xcolor}
\usepackage{etex}
\usepackage{amscd,amsmath,amssymb,amsfonts,xspace,mathrsfs,mathtools,amsthm}
\usepackage[utf8]{inputenc}
\usepackage{bbm}
\usepackage{graphicx}
\usepackage{tabu}
\usepackage[color,matrix,arrow]{xy}
\usepackage{wasysym}
\usepackage{stmaryrd}
\usepackage[shortlabels]{enumitem}
\usepackage{array,amsmath,booktabs} 
\usepackage{slashed}
\usepackage{tensor}
\usepackage{caption,subcaption}
\usepackage{enumitem}
\usepackage[textsize=tiny]{todonotes}
\usepackage{lineno}

\newcommand{\out}{\mathrm{Out}} 
\newcommand{\aut}{\mathrm{Aut}}
 
\newcommand{\soe}{\mathrm{SO}(8)} 
\newcommand{\tra}{\mathsf{T}} 
\newcommand{\frt}{\mathfrak T} 
\newcommand{\scrt}{\mathscr T} 
\newcommand{\tri}{\scrt} 
\newcommand{\scrm}{\mathscr M} 
\newcommand{\bma}{\bfm_\text{A}} 
\newcommand{\bmb}{\bfm_\text{B}} 
\newcommand{\bmc}{\bfm_\text{C}} 
\newcommand{\bmd}{\bfm_\text{D}}

\newcommand{\ua}{u_\text{A}} 
\newcommand{\ub}{u_\text{B}} 
\newcommand{\uc}{u_\text{C}} 
\newcommand{\ud}{u_\text{D}} 
\newcommand{\ue}{u_\text{E}} 
\newcommand{\uf}{u_\text{F}} 
\newcommand{\ug}{u_\text{G}} 
 
\newcommand{\ffc}{f}
\newcommand{\tta}{\text{A}} 
\newcommand{\ttb}{\text{B}} 
\newcommand{\ttc}{\text{C}} 
\newcommand{\ttd}{\text{D}} 
 
\newcommand{\ttf}{\text{F}}

\newcommand{\ttu}{\mathtt{u}} 
\newcommand{\tua}{\mathtt{u}_{\text{A}}}
\newcommand{\tub}{\mathtt{u}_{\text{B}}}
\newcommand{\tuc}{\mathtt{u}_{\text{C}}}
\newcommand{\tud}{\mathtt{u}_{\text{D}}}

\newcommand{\ubp}{u_{\text{bp}}}

\newcommand{\mad}{m_{\text{AD}}}

\newcommand{\ord}{\text{ord}}

\newcommand{\slz}{\text{SL}(2,\mathbb Z)} \newcommand{\slr}{\text{SL}(2,\mathbb R)} 
\newcommand{\spin}{\text{Spin}} 
\newcommand{\tq}{\mathtt{q}} 
\newcommand{\nstar}{\CN=2^*}

\newcommand{\jt}{\vartheta}
\newcommand{\tn}{\tau_0} 
\newcommand{\tuv}{\tau_{\text{\tiny{UV}}}}
\newcommand{\uv}{\tau_{\text{\tiny{UV}}}} 
\newcommand{\luv}{\lambda(\tn)} 
\newcommand{\lau}{\lambda(\tau)}

\newcommand{\be}{\begin{equation}} 
\newcommand{\ee}{\end{equation}} 
\newcommand{\bes}{\begin{equation*}}
\newcommand{\ees}{\end{equation*}}

\newcommand{\im}{i}
  
\newcommand{\CB}{\mathcal{B}}

\newcommand{\CF}{\mathcal{F}}

\newcommand{\CJ}{\mathcal{J}}

\newcommand{\CL}{\mathcal{L}} 
\newcommand{\CM}{\mathcal{M}}  
\newcommand{\CN}{\mathcal{N}}

\newcommand{\CR}{\mathcal{R}}
\newcommand{\CS}{\mathcal{S}}
\newcommand{\CT}{\mathcal{T}}

\newcommand{\BZ}{\mathbb{Z}}

\newcommand{\bra}{\langle}
\newcommand{\ket}{\rangle}

\newcommand{\bfF}{{\boldsymbol F}} 
 
\newcommand{\bfm}{{\boldsymbol m}} 

\newcommand{\bfu}{{\boldsymbol u}}

\newcommand{\bfk}{{\boldsymbol k}}

\newcommand{\SUF}{\text{SU(4)}}

\newcommand{\SUT}{\text{SU(2)}}
\newcommand{\SUO}{\text{U(1)}}
\newcommand{\SL}{\text{SL}(2,\BZ)}

\makeatletter
\newtheorem*{rep@theorem}{\rep@title}
\newcommand{\newreptheorem}[2]{%
\newenvironment{rep#1}[1]{%
 \def\rep@title{#2 \ref{##1}}%
 \begin{rep@theorem}}%
 {\end{rep@theorem}}}
\makeatother

\makeatletter
\newcommand*{\rom}[1]{\expandafter\@slowromancap\romannumeral #1@}
\makeatother

\newtheorem{definition}{Definition}

\newreptheorem{definition}{Definition}

\title{Four flavours, triality and bimodular forms}

\author{Johannes Aspman$^a$, Elias Furrer$^b$, Jan Manschot$^c$ \\
{\it School of Mathematics, Trinity College, Dublin 2, Ireland\\
\it Hamilton Mathematical Institute, Trinity College, Dublin 2 \vspace{20pt}

$^a$\href{mailto:aspmanj@maths.tcd.ie}{aspmanj@maths.tcd.ie}\\
$^b$\href{mailto:furrere@maths.tcd.ie}{furrere@maths.tcd.ie}\\
$^c$\href{mailto:manschot@maths.tcd.ie}{manschot@maths.tcd.ie}

}}
   
\abstract{
We consider $\mathcal{N}=2$ supersymmetric SU(2) gauge theory with $N_f=4$ massive hypermultiplets. The duality group of this theory contains transformations acting on the UV-coupling $\tau_{\text{\tiny{UV}}}$ as well as on the running coupling $\tau$. We establish that subgroups of the duality group act separately on $\tau_{\text{\tiny{UV}}}$ and $\tau$, while a larger group acts simultaneously on $\tau_{\text{\tiny{UV}}}$ and $\tau$. For special choices of the masses, we find that the duality groups can be identified with congruence subgroups of $\text{SL}(2,\mathbb Z)$. We demonstrate that in such cases, the order parameters are instances of bimodular forms with arguments $\tau$ and $\tau_{\text{\tiny{UV}}}$. Since the UV duality group of the theory contains the triality group of outer automorphisms of the flavour symmetry SO(8), the duality action gives rise to an orbit of mass configurations. Consequently, the corresponding order parameters combine to vector-valued bimodular forms with $\text{SL}(2,\mathbb Z)$ acting simultaneously on the two couplings.
}

\preprint{}

\begin{document}
\maketitle

\section{Introduction}
The $\CN=2$ supersymmetric Yang-Mills field theory with gauge group
SU(2) and $N_f=4$ fundamental hypermultiplets is distinguished
for various reasons \cite{Seiberg:1994aj}, including:
\begin{itemize}
\item The theory is superconformal up to mass terms for the hypermultiplets, and is a benchmark for four-dimensional  SCFT's with $\CN = 2$ supersymmetry  \cite{Argyres:1995xn, Argyres:2007tq,
    Argyres:2016xua, Argyres:2020nrr}.
\item The theory is a building block for other four-dimensional $\CN=2$ SCFT's 
  and the 2d/4d correspondence \cite{Gaiotto:2009we, Alday:2009aq}.
\item The theory exhibits an intriguing electric-magnetic duality
group including triality \cite{Seiberg:1994aj}. This duality group acts on the UV coupling
$\tau_{\text{\tiny{UV}}}$ and running coupling constant $\tau$, and contains
elements which act simultaneously on the two couplings as well as separately. 
 \item The theory is a
``parent'' theory from which the asymptotically free $\CN=2$, SU(2)
theories with $N_f\leq 3$ hypermultiplets can be obtained by decoupling one
or more hypermultiplets \cite{Seiberg:1994aj,Seiberg:1994rs}.
\end{itemize}

The focus of the present paper is on the third bullet point. We analyze
duality groups for the couplings $\tau_{\text{\tiny{UV}}}$ and $\tau$
of the $N_f=4$ theory as function of the masses. To this end, explicit expressions for the order
parameter $u=\langle \text{Tr} \, \phi^2 \rangle$, with $\phi$
the complex scalar of the vector multiplet, are determined as function of both $\tau_{\text{\tiny{UV}}}$ and
$\tau$. We identify several loci in the space of
masses where $u$ transforms as a modular form for both $\tau_{\text{\tiny{UV}}}$ and $\tau$. This
extends our recent work \cite{aspman2021cutting} on theories with
$N_f\leq 3$ to $N_f=4$. In \cite{aspman2021cutting}, we 
determined fundamental domains for the running coupling $\tau$ for the asymptotically free
theories by analyzing in detail the
order parameter $u$ as function of the
effective coupling $\tau\in \mathbb H$. We have demonstrated that for generic
masses this function has branch points, with the consequence that the
fundamental domain for $\tau$ is in general not of the form $\Gamma\backslash\mathbb H$ for a congruence
subgroup $\Gamma\subset \slz$. Only for specific values of the
masses, such as those giving rise to  Argyres-Douglas (AD) points, the order parameter is (weakly)
holomorphic as function of $\tau$, and the fundamental domain is that
of a congruence subgroup. In this paper, we find that these
features are present as well for the $N_f=4$ theory, but with
an additional dependence on $\tau_{\text{\tiny{UV}}}$.

At  special modular loci, some properties of the $N_f=4$ order parameters are similar to that of
the $\nstar$ SU(2) theory, i.e. the superconformal theory with a single
adjoint hypermultiplet. The $\nstar$ order parameter
transforms as a modular form under the group
$\Gamma(2)\times \Gamma(2)$ with the first factor acting on $\tau$ and the second
on $\tuv$, while it also transforms as a modular form under simultaneous
$\slz$ transformations of $\tau$ and $\tuv$ \cite{Ferrari:1997gu,
  Labastida:1998sk, Huang:2011qx}. It was
later clarified that $u(\tau,\tuv)$ is an
example of a meromorphic \emph{bimodular} form \cite{Manschot:2021qqe}. Such functions have appeared, although sporadically, in the mathematical literature 
\cite{stienstra_zagier_16,yang2007differential,liuquan2020}. 

The $\SUT$ $N_f=4$ theory exhibits a richer structure: It has four mass
parameters that give rise to seven singular vacua on the
$u$-plane. For special choices of the masses, the $u$-planes contain
any of the three $\SUT$ Argyres-Douglas superconformal points
$(A_1,A_2)$, $(A_1,A_3)$ and $(A_1, D_4)$, while in the massless
case there is a non-abelian Coulomb point with a five
quaternionic-dimensional Higgs branch
\cite{Argyres:1995jj,Argyres:1995xn}.  
 For generic masses, the singularities are roots of a sextic
 polynomial, for which there is no known expression. The flavour
 symmetry $\soe$ becomes the universal cover $\spin(8)$ in the quantum
 theory. It has a \emph{triality} group $\out(\spin(8))$ of outer
 automorphisms, which is isomorphic to the symmetric group $ S_3$ on
 three letters. The full symmetry group of the $N_f=4$ curve is then
 the semidirect product $\spin(8)\rtimes_\varphi \slz$, which is
 induced by the group homomorphism $\varphi: \slz\to
 \out(\spin(8))$. As the triality group is of order $| S_3|=6$, the
 orbits of the group action on mass space $\mathbb C^4$ generally have
 six elements. However, there are specific mass configurations with
 enhanced global symmetry that are invariant under subgroups of the
 triality group, for which the orbits collapse, either to three
 elements or to a single element.

We study four such configurations in detail, and show that their order parameters, periods and discriminants are bimodular forms  for subgroups of $\slz$. For the triality invariant  case $(m_1,m_2,m_3,m_4)=(m,m,0,0)$ we find that the order parameter is a bimodular form of weight $(0,2)$ with $\Gamma(2)$ acting on both $\tau$ and $\tuv$ individually, while it is also bimodular for $\slz$ acting on $\tau$ and $\tuv$ simultaneously. If all four hypermultiplets are rather given an equal mass, the triality orbit has three elements. The $u$-planes for these three mass configurations are modular curves for the three subgroups of $\slz$ conjugate to $\Gamma^0(4)$. The order parameters, periods and discriminants are permuted by triality, and can thus be organised into vectors to form  one-parameter families of vector-valued bimodular forms for $\slz$. We further give some examples of exact expressions for order parameters of more complicated theories with two independent mass parameters. These theories then include both AD points and branch cuts.  

The outline of the paper is as follows. In Section  \ref{sec:nf4} we discuss the symmetries of the $N_f=4$ SW curve, and study the action of the triality group on the mass space. Specific mass configurations with enhanced global symmetry are then studied in Section \ref{SWcurve}, where we also provide a definition of bimodular forms and vector-valued bimodular forms most suited for our analyses.
The Appendix \ref{app:modularforms} contains relevant properties of congruence subgroups and modular forms. In Appendix \ref{app:partitions}, we study the possible singularity spectrum of the $N_f=4$ theory. In Appendix  \ref{sec:qqcurve} we obtain similar results for the $N_f=4$ curve constructed from the qq-characters of the $N_f=4$ theory. Appendix \ref{sec:masslessnf0123} finally provides expressions for the limits to the asymptotically free theories.

\section{Four flavours and triality}\label{sec:nf4}
The one-loop beta function of $\CN=2$ supersymmetric Yang-Mills theory with $N_f\leq 4$ hypermultiplets in the fundamental representation  is $\beta_{N_f}(g_{\text{YM}})=-\frac{g_{\text{YM}}^3}{16\pi^2}(4-N_f)$. The gauge coupling $g_{\text{YM}}$ is combined with the theta angle $\theta$ in the Lagrangian as  $\tau=\frac{\theta}{\pi}+\frac{8\pi\im}{g_{\text{YM}}^2}$. This complexified gauge coupling can be considered as the expectation value of a background chiral superfield. In the renormalisation scheme where the superpotential remains a holomorphic function of all chiral superfields, the one-loop running coupling at the energy scale $E$ can be expressed as \cite{Intriligator:1995au}
\begin{equation}
\tau(E)=\tuv-\frac{4-N_f}{2\pi\im}\log \frac{E}{\Lambda_{\text{UV}}}.
\end{equation}
It is one-loop exact in the holomorphic scheme, and thus for $N_f<4$ the combination
\begin{equation}
\Lambda_{N_f}^{4-N_f}\coloneqq \Lambda_{\text{UV}}^{4-N_f}e^{2\pi\im\tuv}
\end{equation}
of the scale $\Lambda_{\text{UV}}$ and the coupling $\tuv$  is invariant to all orders in perturbation theory. This complexified dynamical scale  $\Lambda_{N_f}$ sets the overall scale of the theory. For $N_f=4$ on the other hand,  there is a distinguished dimensionless parameter $\tuv$, on which the theory depends nontrivially. To shorten the notation, we will also set $\tn:=\tuv$ and $q_0:=e^{2\pi\im\tn}$ in the following.

\subsection{The curve}\label{sec:curve}
The low-energy physics of $\CN=2$ SYM with $N_f=4$ massive hypermultiplets has been determined in \cite{Seiberg:1994aj,Minahan:1996ws,Ferrari:1997gu,Dorey:1996bn,Argyres:1999ty,Huang:2011qx}. 
Similar to the asymptotically free ($N_f\leq 3$) cases, the physics is
encoded in an elliptic curve which depends holomorphically on the
Coulomb branch parameter $u\in \CB_4$. This coordinate $u$
parametrises the Coulomb branch $\CB_4$ of the $N_f=4$ theory. Let us
first define the symmetric mass combinations
\begin{equation}\begin{aligned} \label{bracketnotation}
& \left\llbracket m_{1}^{k}\right\rrbracket =\sum_{i=1}^{4}m_{i}^{k},
\qquad \quad \,\, \left\llbracket m_{1}^{2}m_{2}^{2}\right\rrbracket =\sum_{i<j}m_{i}^{2}m_{j}^{2}\\
&\left\llbracket m_{1}^{4}m_{2}^{2}\right\rrbracket =\sum_{i\neq
  j}m_{i}^{4}m_{j}^{2}, \quad \left\llbracket
  m_{1}^{2}m_{2}^{2}m_3^2\right\rrbracket
=\sum_{i<j<k}m_{i}^{2}m_{j}^{2}m_k^2,\\
&\mathrm{Pf}(\bfm)=m_{1}m_{2}m_{3}m_{4}. 
\end{aligned}\end{equation}
The $N_f=4$ curve for generic masses is then \cite{Seiberg:1994aj}
\begin{equation}\label{nf4genericcurve}
y^{2} = W_{1}W_{2}W_{3}+A\left(W_{1}T_{1}\left(e_{2}-e_{3}\right)+W_{2}T_{2}\left(e_{3}-e_{1}\right)+W_{3}T_{3}\left(e_{1}-e_{2}\right)\right)-A^{2}N,
\end{equation}
where
\begin{eqnarray}
W_{i} & = & x-e_{i}u-e_{i}^{2}R,\nonumber \\
A & = & \left(e_{1}-e_{2}\right)\left(e_{2}-e_{3}\right)\left(e_{3}-e_{1}\right),\nonumber \\
R & = & \frac{1}{2}\left\llbracket {m}_{1}^{2}\right\rrbracket ,\nonumber \\
T_{1} & = & \frac{1}{12}\left\llbracket {m}_{1}^{2}{m}_{2}^{2}\right\rrbracket -\frac{1}{24}\left\llbracket {m}_{1}^{4}\right\rrbracket ,\nonumber \\
T_{2,3} & = & \mp\frac{1}{2}\mathrm{Pf}(\bfm)-\frac{1}{24}\left\llbracket {m}_{1}^{2}{m}_{2}^{2}\right\rrbracket +\frac{1}{48}\left\llbracket {m}_{1}^{4}\right\rrbracket ,\nonumber \\
N & = & \frac{3}{16}\left\llbracket {m}_{1}^{2}{m}_{2}^{2}{m}_{3}^{2}\right\rrbracket -\frac{1}{96}\left\llbracket {m}_{1}^{4}{m}_{2}^{2}\right\rrbracket +\frac{1}{96}\left\llbracket {m}_{1}^{6}\right\rrbracket,\label{eq:SWfund1}
\end{eqnarray}
and the half periods
\begin{equation}\label{halfperiods}
e_1=\frac13(\vartheta_3^4+\vartheta_4^4),\quad e_2=-\frac13(\vartheta_2^4+\vartheta_3^4)\quad e_3=\frac13(\vartheta_2^4-\vartheta_4^4)
\end{equation}
are functions of $\tn=\tuv$, with $e_1+e_2+e_3=0$. The Jacobi theta functions $\jt_i$ are defined in Appendix \ref{app:modularforms}. Since the rhs of \eqref{nf4genericcurve} is a cubic polynomial in $x$, it is indeed an elliptic curve. 
We obtain the low energy theory with $N_f=3$ flavours by taking the limit $\tn\to\im\infty$ (or, equivalently, $q_0\to 0$) and $m_4\to \infty$ while holding $\Lambda_3=64q_0^{\frac12}m_4$ fixed.  The order parameters are then related as \cite{Seiberg:1994aj}
\begin{equation}\label{nfdecoupling}
u_{N_f=4}+\frac14 e_1 \left\llbracket m_{1}^{2}\right\rrbracket\to u_{N_f=3}.
\end{equation}
 See Appendix \ref{sec:masslessnf0123} for the corresponding curves.

Let us study the singularity structure of the Coulomb branch. For generic masses $\bfm=(m_1,m_2,m_3,m_4)$, there are six distinct strong coupling singularities. By tuning the mass, some of those singularities can collide.  If we weight each singularity by the number of massless hypermultiplets at that point, the total weighted number of singularities on the $u$-plane is thus always $6$. Denote by $k_l$ the weight of the $l$-th singularity, and by $\bfk(\bfm)=(k_1,k_2,\dots)$ the vector of those weights. In Table \ref{masscases}, we list a selection of specifically symmetric mass configurations. One notices that certain a priori unrelated cases have the same weight vector $\bfk$ and global symmetries, such as the cases $\{\text{B, C, D}\}$ and $\{\text{E, F, G}\}$.  This will be explained in the next subsection. It is also clear that $\bfk(\bfm)$ gives a 
partition of $6$, the total number of singularities on $\CB_4$. 
As there are $p(6)=11$ such partitions, it is a natural question whether all of those 11 partitions are realised as $\bfk(\bfm)$ for a mass $\bfm$. We study this question in Appendix \ref{app:partitions}.

\begin{table}[h]\begin{center}
$\begin{tabu}{| c|c|c|c|c |} 
\hline 
\text{Name} & \bfm& \bfk(\bfm)&\text{global symmetry} \\ \hline
\text{A}&(m,m,0,0)&(2,2,2)&\SUT\times \SUT\times \SUT\times \SUO\\
\text{B}&(m,m,m,m)& (4,1,1)&\SUF\times \SUO \\
\text{C}&(2m,0,0,0)& (4,1,1)&\SUF\times \SUO \\
\text{D}&(m,m,m,-m)&(4,1,1)&\SUF\times \SUO\\
\text{E}&(m,m,\mu,\mu)&(2,2,1,1)&\SUT\times \SUT\times \SUO\times \SUO\\
\text{F}&(m+\mu,m-\mu,0,0)&(2,2,1,1)&\SUT\times \SUT\times \SUO\times \SUO\\
\text{G}&(m,m,\mu,-\mu)&(2,2,1,1)&\SUT\times \SUT\times \SUO\times \SUO\\
\hline 
\end{tabu}$
\caption{List of some mass cases with enhanced flavour symmetry in $N_f=4$, with $\mu\neq m$. The vector $\bfk(\bfm)$ lists the multiplicities of all singularities on the Coulomb branch $\CB_4$ with mass $\bfm$.}\label{masscases}\end{center}
\end{table}

\subsection{Triality}\label{sec:nf4massive}
Let us study the symmetries of the $N_f=4$ curve \eqref{nf4genericcurve} with mass  $\bfm=(m_1,m_2,m_3,m_4)$. Scale invariance, the $\SUO_R$ R-symmetry and the $\slz$ symmetry acting on $\tn$ are explicitly broken by the masses. There is a remnant scale invariance on the Coulomb branch, which manifests itself in the $\CJ$-invariant $\CB_4\times \mathbb C^4\times \mathbb H\to\mathbb C$ of the curve  being a quasi-homogeneous rational function of degree 0 and type $(2,1,0)$, 
\begin{equation}\label{scalingsymmetry}
\CJ(s^2 u,s \,\bfm,\tn)=\CJ(u,\bfm,\tn), \quad s\in \mathbb C^*.
\end{equation}
The $N_f=4$ theory has an $\soe$ flavour symmetry, which becomes the universal double cover $\spin(8)$ in the quantum theory. In particular, there exists a short exact sequence
\begin{equation}
1\to \mathbb Z_2\to \spin(8)\to \soe\to 1
\end{equation}
of Lie groups. 
The cover $\spin(8)$ has an order 6 group $\out(\spin(8))$ of outer automorphisms, which is isomorphic to $S_3$ \cite{Adams:623418,Fulton2004}.\footnote{For any Lie group $G$, there are three associated groups. $\text{Aut}(G)$ is the Lie group consisting of all automorphisms of $G$  (i.e. group isomorphisms $G\to G$), $\text{Inn}(G)$ is a normal subgroup of $\text{Aut}(G)$ consisting of inner automorphisms given by $\alpha_g(h)\coloneqq ghg^{-1}$ for any $g\in G$, and $\text{Out}(G)=\text{Aut}(G)/\text{Inn}(G)$ is the quotient group. The automorphism group of $\spin(8)$ is $\aut(\mathrm{SO}(8))=\text{PSO}(8)\rtimes S_3$ \cite{Adams:623418}.}

\begin{figure}[h]\centering
	\includegraphics[scale=2]{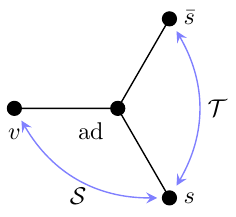}
	\caption{Dynkin diagram of $\mathfrak d_4=\mathrm{Lie}(\spin(8))$. The group $\mathscr T\cong S_3$ of outer isomorphisms acts by permutations on the three conjugacy classes of irreducible representations $v$, $s$ and $\bar s$ attached to the nodes of the diagram. The 28-dimensional adjoint representation is left invariant by $\mathscr T$.}\label{fig:d4}
\end{figure}

This group of outer automorphisms acts on the $N_f=4$ theory as follows. The states with $(n_m,n_e)=(0,1)$ are the elementary hypermultiplets, which transform in the fundamental vector representation of $\spin(8)$. The magnetic monopole $(1,0)$ transforms as one spinor representation, and the dyon $(1,1)$ transforms as the conjugate spinor representation \cite{Seiberg:1994aj}. By an accidental isomorphism, these three representations are all $8$-dimensional and irreducible, and they are permuted by the outer automorphism group $\out(\spin(8))\cong S_3$. It is generated by 
\begin{equation}\label{STtriality}
\CT=\begin{pmatrix}1&0&0&0\\ 0&1&0&0\\0&0&1&0\\0&0&0&-1\end{pmatrix}, \qquad \CS=\frac12\begin{pmatrix} 1&1&1&1\\ 1&1&-1&-1\\ 1&-1&1&-1\\ 1&-1&-1&1\end{pmatrix},
\end{equation}
which act on the column vector $\bfm\in\mathscr M\coloneqq \mathbb C^4$ from the left \cite{Seiberg:1994aj,Marino:1998ru,Tai:2010im}.
The map $\CT$ exchanges the two spinors keeping the vector fixed, while $\CS$ exchanges the vector with the spinor, keeping the conjugate spinor fixed. This is depicted in Fig. \ref{fig:d4}.

The generators \eqref{STtriality} satisfy the algebra 
\begin{equation}\label{s3triality}
\CT^2=\CS^2=(\CS\CT)^3=\CS\CT^2\CS=\mathbbm 1,
\end{equation}
which is a presentation of the symmetric group $S_3$. Since  $\CT^\mathsf{T}\CT=\CS^\mathsf{T}\CS=\mathbbm 1$ but $\det \CT=\det\CS=-1$, the  matrices $\CT$ and $\CS$ generate a subgroup
\begin{equation}\label{scrT}
\mathscr T=\langle \CT,\CS\rangle 
\end{equation}
of the orthogonal group $\text{O}(4,\mathbb C)$, isomorphic to $S_3$.\footnote{They actually form a subgroup of $O(4,\mathbb Q)$, but  act on $\bfm\in \mathbb C^4$.} As a consequence, they leave the inner product $\left\llbracket m_{1}^{2}\right\rrbracket$ \eqref{bracketnotation} invariant.

\begin{figure}[h]\centering
	\includegraphics[scale=1]{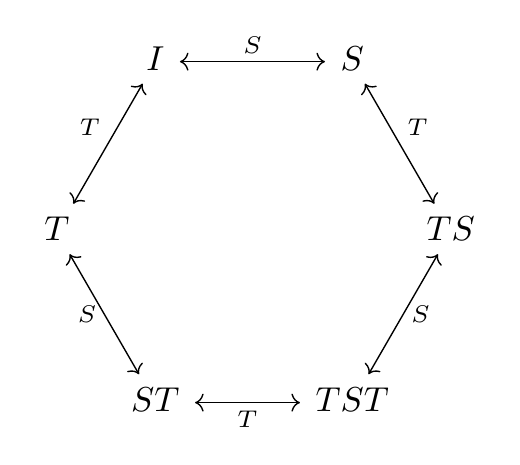}
	\caption{Action of $\slz$ on $\slz/\Gamma(2)\cong S_3$}\label{fig:gamma2}
\end{figure}

The flavour symmetry mixes with the $\slz$-symmetry acting on the UV-coupling $\tn$ in an interesting way. To see this, notice that the reduction $\mathbb Z\to \mathbb Z_2$ modulo $2$ induces a  homomorphism  $\slz\to \text{SL}(2,\mathbb Z/2\mathbb Z)$. 
Since $\text{SL}(2,\mathbb Z/2\mathbb Z)\cong  S_3$ are isomorphic, by transitivity we have a group homomorphism 
\begin{equation}\label{phihomomorphism}
\varphi:  \slz\longrightarrow \out(\spin(8)).
\end{equation}
The full symmetry group of the $N_f=4$ theory is the semidirect product \cite{Seiberg:1994aj}\footnote{Recall that for two groups $G$ and $H$, a group homomorphism $\varphi: G\to \aut(H)$ defines a semi-direct product $H\rtimes_\varphi G\subset H\times G$ with the multiplication $(h_1,g_1)(h_2,g_2)\coloneqq (h_1\varphi(g_1)(h_2),g_1g_2)$. For $(h,g)\in H\rtimes_\varphi G$, the inverse is found as $(\varphi(g^{-1})(h^{-1}),g^{-1})$.}
\begin{equation}\label{frt}
\frt\coloneqq \spin(8)\rtimes_\varphi \slz.
\end{equation}
The group $(\frt,\bullet)$ consists of elements $(A,\gamma)\in \spin(8)\times \slz$, with group operation
\begin{equation}
(A,\gamma)\bullet (\tilde A,\tilde \gamma)\coloneqq (A\,  \varphi(\gamma)(\tilde A), \gamma\,\circ\,\tilde\gamma).
\end{equation}
The action of \eqref{STtriality} is thus accompanied with an action of $\slz$ on $\tau$ and $\tn$. From \eqref{s3triality} we find that $\CT^2$ and $\CS\CT^2\CS$ leave any mass configuration invariant. This implies that the theory should also be invariant under the simultaneous action of $T^2$ and $ST^2S$ on the two couplings. These two matrices in $\slz$ generate the principal congruence subgroup $\Gamma(2)$. From this it is also clear that 
\begin{equation}\label{sl2zmodgamma2}
\slz/\Gamma(2)=\{I,T,S,TS,ST,TST\}\cong S_3,
\end{equation}
which is another way to see that the group of outer isomorphisms is $S_3$ \cite{Tai:2010im}. This action is depicted in Fig. \ref{fig:gamma2}. 
The subgroup $\Gamma(2)$ is the kernel of the above group homomorphism $\slz\to \text{SL}(2,\mathbb Z/2\mathbb Z)$, such that it is in fact a normal subgroup $\Gamma(2)\lhd\,\slz$.

\begin{figure}[h]\centering
	\includegraphics[scale=1.3]{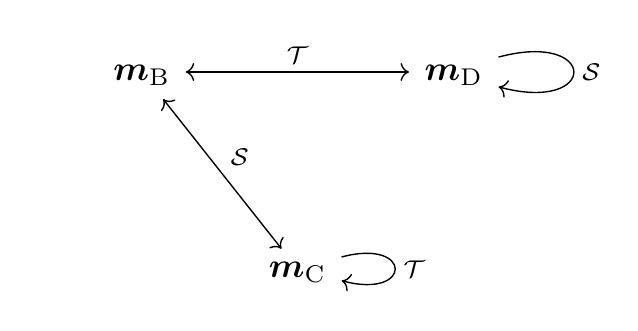}
	\caption{Relation among $\bfm_\ttb=(m,m,m,m)$, $\bfm_\ttc=(2m,0,0,0)$ and  $\bfm_{\text{D}}=(m,m,m,-m)$. }\label{fig:BCD}
\end{figure}

The moduli spaces of the cases A--G of Table \ref{masscases} are related by $\mathscr T$ in the following way. We have that $\bfm_\tta, \bfm_\ttc,\bfm_\ttf$ are invariant under $\CT$. Case A is invariant under both $\CT$ and $\CS$. The $\CS$-transformation relates cases B and C, as well as E and F, while leaving cases D and G invariant. 
We depict the relation among cases B, C and D in Fig. \ref{fig:BCD}. For the cases E, F and G, there is an analogous diagram.
An instance of these relations is that the weights of the singular structure on the Coulomb branch are invariant under those spaces that are related by triality, 
\begin{equation}
\bfk(\mathscr T \bfm)=\bfk(\bfm).
\end{equation}

\begin{figure}[h]\centering
	\includegraphics[scale=1]{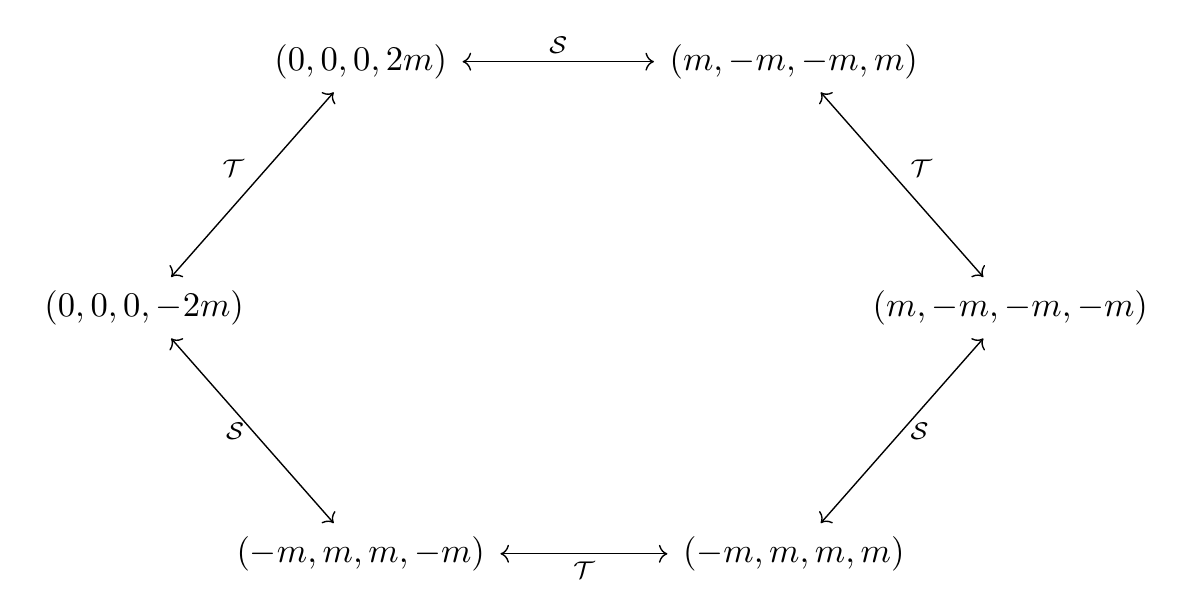}
	\caption{Orbit of the mass vector $\bfm=(0,0,0,2m)$ under $\CT$ and $\CS$. }\label{fig:m4}
\end{figure}

Using the action of the SO(8) flavour group, a large range of masses with
equivalent duality diagrams can
be reached. For example, the mass $\bfm=(2m,0,0,0)$ is related to
$\bfm=(0,0,0,2m)$ by an SO(8) rotation. The first one is invariant under
$\CT$ while the second one is not. The orbit under $\CT$ and $\CS$ for the case $\bfm=\bfm_{\text{B}}=(2m,0,0,0)$
is, as we have just discussed, given by Fig. \ref{fig:BCD}, while that
of $\bfm=(0,0,0,2m)$ is given in Fig. \ref{fig:m4}. We see that it is
of order six, and includes different relative signs compared to 
$\bma$ and $\bmd$. On closer inspection, we note that the mass vectors come in pairs differing by an
overall sign, which is an element of SO(8). Thus identifying the mass
vectors related by SO(8) in diagram \ref{fig:m4}, we find that it is
equivalent to diagram \ref{fig:BCD}.

\subsection{Group action}\label{sec:groupaction}
The action 
\begin{equation}\begin{aligned}\label{groupaction}
\mathscr T\times \scrm&\longrightarrow \scrm \\
(g,\bfm)&\longmapsto g\cdot\bfm
\end{aligned}\end{equation}
of the triality group $\mathscr T$ on mass space $\scrm$ can be studied in great detail. It is easy to check that the action is faithful\footnote{For every $g\neq h\in \mathscr T$ there exists an $\bfm\in\scrm$ such that $g\cdot\bfm\neq h\cdot\bfm$.}, but neither free\footnote{A group action is free if it has no fixed points, but $\bfm=0$ is a fixed point for any $g\in\mathscr T$.} nor transitive\footnote{For each pair $\bfm, \tilde\bfm\in\scrm$ there exists $g\in\mathscr T$ such that a $g\cdot\bfm=\tilde\bfm$. A counterexample would be $\bfm=0$ and $\tilde\bfm\neq 0$.}.

Up to conjugation, $S_3\cong\mathscr T$ has four subgroups. They are: the trivial group $\mathbb Z_1$, the symmetric group $S_2\cong\mathbb Z_2$, the alternating group $A_3\cong\mathbb Z_3$, and $S_3$ itself. They have order 1, 2, 3, and 6, respectively. All three proper subgroups are abelian. For a given $\bfm$, triality thus not always acts by the full $S_3$ but rather by a subgroup. For every $\bfm\in\scrm$ we can study the orbit $\tri\cdot\bfm=\{g\cdot\bfm \, |\, g\in\mathscr T\}$. The sets of orbits of $\scrm$ then give a partition of $\scrm$ under the action \eqref{groupaction}.

First, notice that since $\tri$ is a finite group, all elements have finite order. In particular, $\CT^2=\CS^2=(\CT\CS\CT)^2=\mathbbm1$ and $(\CS\CT)^3=(\CT\CS)^3=\mathbbm 1$. The stabiliser subgroup of a mass $\bfm\in\scrm$ is defined as $\tri_{\bfm}=\{g\in\tri \,|\, g\cdot \bfm=\bfm\}$. By the orbit-stabiliser theorem
\begin{equation}\label{OStheorem}
|\tri\cdot\bfm|=|\tri|/|\tri_{\bfm}|,
\end{equation}
it suffices to study the fixed point equations in order to identify the stabiliser subgroups $\{\mathbb Z_1, S_2,A_3,S_3\}$ with the subgroups of $\tri$. It is straightforward to identify the fixed point loci
\begin{equation}\label{STloci}
\begin{split}
\CL_{\CT}&=\{\bfm\in\scrm\,|\, m_4=0\}, \\
\CL_{\CS}&=\{\bfm\in\scrm\,|\, m_1=m_2+m_3+ m_4\}, \\
\CL_{\CS\CT\CS}&=\{\bfm\in\scrm\,|\, m_1=m_2+m_3- m_4\}, \\
\CL_{\CS\CT}=\CL_{\CT\CS}&=\{\bfm\in\scrm\,|\, m_1=m_2+m_3\text{ and } m_4=0\}, \\
\end{split}
\end{equation}
where $\CL_g=\{\bfm\in\scrm\,|\, g\cdot\bfm=\bfm\}$. For $\bfm$ in precisely one of $\CL_{\CT}$, $\CL_{\CS}$ or $\CL_{\CS\CT\CS}$, one finds that $|\tri\cdot\bfm|=3$. From \eqref{OStheorem} it then follows that   $|\tri_{\bfm}|=2$, such that $\tri_{\bfm}\cong S_2$. In fact, since $\CT$, $\CS$ and $\CS\CT\CS$ are all order $2$ elements of $\tri$, the stabiliser groups $\tri_{\bfm}$ for $\bfm$ in either of the three loci are precisely the three order 2 conjugate subgroups of $\tri\cong S_3$.\footnote{If we represent $S_3$ in cycle notation of permutations of $\{1,2,3\}$, the three order 2 conjugate subgroups of $S_3$ are $\{(),(1,2)\}$, $\{(),(1,3)\}$ and $\{(),(2,3)\}$.}

The intersection 
\begin{equation}
\CL_1=\CL_{\CT}\cap \CL_{\CS}=\{\bfm\in\scrm\,|\, m_1=m_2+m_3 \text{ and } m_4=0\}
\end{equation}
is the locus of triality invariant masses, $\tri
\cdot\bfm=\bfm$. Thus, according to \eqref{OStheorem} we have
$|\tri_{\bfm}|=6$ for such masses, such that indeed
$\tri_{\bfm}=\tri$.  For the last locus in \eqref{STloci}, we see
immediately that $\CL_{\CS\CT}=\CL_{\CT\CS}=\CL_{\CT}\cap \CL_{\CS}$
contains precisely the invariant masses. Therefore, if $\bfm$ is kept
fixed by either $\CT\CS$ or $\CS\CT$ then it is also fixed by both
$\CT$ and $\CS$ and therefore by all of $\tri$. Since $\CS\CT$ and
$\CT\CS$ are the only elements of $\mathscr T$ of order 3, there is actually no mass $\bfm$ such that $\tri\cdot\bfm$ has 2 elements, and so there is no stabiliser subgroup isomorphic to $A_3$. By case analysis, it is also easy to prove that the set $\tri\cdot\bfm$ has $1$, $3$ or $6$ elements.

\begin{figure}[h]\centering
	\includegraphics[scale=1.2]{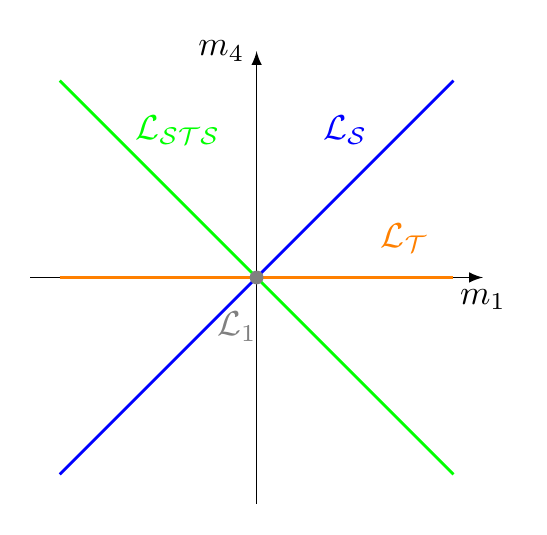}
	\caption{The loci \eqref{STloci} with nontrivial stabiliser groups on the subspace $m_2=m_3=0$ in $\scrm$.    They all mutually intersect in the locus $\CL_1$ of triality invariant masses.}\label{fig:L3}
\end{figure}

Let us summarise. If $\bfm\in \CL_1$, it is invariant under $\tri$. If $\bfm$ is in any of $\CL_{\CT}$, $\CL_{\CS}$ or $\CL_{\CS\CT\CS}$, it could be in the intersection of any two of them. These intersections are however all equal to $\CL_1$, which is of course because any two elements of $\{\CT,\CS,\CS\CT\CS\}$ generate $\tri$. This is depicted in Fig. \ref{fig:L3}.

If $\bfm$ is then an element of 
\begin{equation}
\CL_3=\CL_{\CT}\cup \CL_{\CS}\cup  \CL_{\CS\CT\CS} \setminus \CL_1,
\end{equation}
then the stabiliser group of $\bfm$ is isomorphic to $S_2$. If $\bfm$ does not lie in either $\CL_1$ or $\CL_3$, then there is no remaining symmetry. It lies in 
\begin{equation}
\CL_6=\scrm\setminus \CL_1\cup\CL_3,
\end{equation}
and its stabiliser group is trivial.

\section{Order parameters and bimodular forms}\label{SWcurve}
For the $N_f=4$ SW theory, there are  other curves than the one introduced by Seiberg and Witten \cite{Huang:2011qx,Dorey:1996bn,Nekrasov:2015wsu,Jeong:2019fgx,Manschot:2019pog,Nekrasov:2012xe,Gaiotto:2009hg,Gaiotto:2009we,Witten:1997sc}. In this paper, we focus on the modularity of the original SW curve \eqref{nf4genericcurve}. In Appendix \ref{sec:qqcurve} we show that similar results hold for the curve constructed from the qq-characters of the theory. 
The $N_f=4$ SW curve however has the advantage over the qq-curve in that it depends explicitly on $\tn$, and one can study modular transformations of $\tn$. We proceed by studying the mass configurations with the largest flavour symmetry groups, A, B, C and D. In all these cases, $u$ is a weight $(0,2)$ bimodular form, which we define below, for a triple of groups related to the duality group of the decoupling theory where the mass of the hypermultiplets is  infinitely large, and the stabiliser group of the mass under the triality action. Since case A is triality invariant, $u$ in that case transforms under the full $\slz$ group. The other cases B, C and D are permuted by triality, and furnish a vector-valued bimodular form.

The massless case where $\bfm_0=(0,0,0,0)$ is very simple, as $j(\tau)=\CJ(u,0,\tn)=j(\tn)$ and therefore 
\begin{equation}\label{0000sol}
\tau(u)=\tn
\end{equation}
is constant over the whole Coulomb branch $\CB_4\ni u$. In other words, the coupling $\tau$ is fixed and thus does not run, which is a consequence of the massless $N_f=4$ theory being exactly superconformal. There are six singularities, which all sit at the origin $u=0$ and form the non-abelian Coulomb point with a five quaternionic-dimensional Higgs branch \cite{Argyres:1995xn}.

Let us recall a method for finding explicit expressions for the Coulomb branch parameter $u$. This was recently discussed in  detail in \cite{aspman2021cutting}. The $\CJ$-invariant of the SW curve \eqref{nf4genericcurve} is a rational function $\CJ(u,\bfm,\tn)$ in $u$, the masses $\bfm=(m_1,m_2,m_3,m_4)$ and $e_i(\tn)$. While in general it is not possible to solve $\CJ(u(\tau),\bfm,\tn)=j(\tau)$ for $u(\tau)$ analytically, for specific masses we can rather solve $\CJ(u,\bfm,\tn)=\CR(\lambda)$ for $u$, where 
\begin{equation}\label{cjlambda}
\CR(p)=2^8\frac{(1+(p-1)p)^3}{(p-1)^2p^2}
\end{equation}
is the unique rational function with the property that $\CR(\lambda)=j$. Here,  $\lambda\coloneqq \frac{\jt_2^4}{\jt_3^4}$ is the modular lambda function (a Hauptmodul for $\Gamma(2)$) with $\vartheta_i$ the Jacobi theta functions \eqref{Jacobitheta}, and $j$ is the modular $j$-function \eqref{je4e6} . The reason for this is that in certain mass configurations, the sextic equation constructed from the rational function $\CJ(u,\bfm,\tn)$ defines a field extension of $\mathbb C(\slz)$ with intermediate field $\mathbb C(\Gamma(2))$, such that the sextic equation factors over $\mathbb C(\Gamma(2))$ into products of lower degree polynomials \cite{aspman2021cutting}.\footnote{It is true in $N_f=0,1,2,3,4$ that $g_k$ is a polynomial in $u$ of degree $k$, such that  $(g_2^3-27g_3^2)j-12^3g_2^3$ is indeed a sextic polynomial in $u$.}

\subsection{Case A}\label{sec:caseASWcurve}

For the mass $\bma=(m,m,0,0)$, this allows to express $u$ as a rational function in Jacobi theta functions of $\tn$ and $\tau$. There are in fact six solutions to the correspondence $\CJ(u(\tau),\bfm,\tn)=j(\tau)$. A consistent way of choosing which solution to use, which we will employ throughout, is to take the one that has the right decoupling limit when decoupling the massive hypermultiplets, i.e., the one that decouples to the order parameter of massless $N_f=2$, Eq. \eqref{unf2m=0}.

In view of the more complicated mass cases, we can further simplify the rather lengthy expression. The dependence on $\tau$ is in fact only through $\lambda= \frac{\jt_2^4}{\jt_3^4}$. This is not quite true for $\tn$, for which  $u$ has weight 2 \cite{Huang:2011qx}. This weight factor can be extracted by eliminating $\jt_4(\tn)$ through the Jacobi identity \eqref{jacobiabstruseidentity} and $\jt_2(\tn)$ through the definition of $\luv$.  This gives
\begin{equation}\label{uASW}
\ua(\tau,\tn)=-\frac{m^2}{3}\jt_3(\tn)^4\frac{\luv^2+2\left(\lau-1\right)\luv-\lau}{\luv-\lau}.
\end{equation}
The simple mass dependence of $\ua$ is a consequence of the scaling symmetry \eqref{scalingsymmetry}. The second prefactor  $\jt_3(\tn)^4$ gives the weight 2. The remaining quotient is written in a manifestly invariant fashion. Let us denote by $\Gamma_\tau$ ($\Gamma_{\tn}$) a group acting by linear fractional transformations on $\tau$ ($\tn$). As $\jt_3(\tn)^4$ is a modular form of weight $2$ and $\lambda(\tn)$ a modular function (of weight 0) for $\Gamma(2)_{\tn}$, one can easily see that $\ua(\tau,\tn)$ is a weight $2$ modular form for $\Gamma(2)_{\tn}$ for fixed $\tau$, and a modular function for  $\Gamma(2)_{\tau}$ for fixed $\tn$. 
We thus have that
\begin{equation}\label{uagamma2}
\ua(\gamma_1 \tau,\gamma_2 \tn)=(c_2 \tn+d_2)^2 \ua(\tau,\tn), \quad \gamma_i=\begin{pmatrix}a_i&b_i\\c_i&d_i\end{pmatrix}\in\Gamma(2)
\end{equation}
for $i=1,2$. We call $\ua$ modular for $\Gamma(2)_\tau\times \Gamma(2)_{\tn}$, where the occurrence of two groups indicates that they act on both variables $\tau$ and $\tn$ separately.

The mass $\bma$ is invariant under the triality group \eqref{scrT}. As triality acts on $\tau$ and $\tn$ together, this suggests that $\ua$ transforms under a simultaneous transformation of $\slz$. Indeed, if one acts simultaneously on $\tau$ and $\tn$ with $\slz$, it is easy to check from $T: \lambda\mapsto \frac{\lambda}{\lambda-1}$ and $S:\lambda\mapsto 1-\lambda$ that $\ua(\tau,\tn)$ transforms as 
\begin{equation}\label{uasl2z}
\ua(\gamma\tau,\gamma\tn)=(c\tn+d)^2\,\ua(\tau,\tn), \quad \gamma=\begin{pmatrix}a&b\\c&d\end{pmatrix}\in\slz.
\end{equation}
We call $\ua$ modular for $\slz_{(\tau,\tn)}$, where the notation indicates that the single group $\slz$ acts on both $\tau$ and $\tn$ simultaneously. 
The two transformations \eqref{uagamma2} and \eqref{uasl2z}  are 
characteristic properties for  functions known as
``bimodular forms''
\cite{stienstra_zagier_16,yang2007differential,liuquan2020}. For our
application to $N_f=4$ SQCD, we will adopt the following definition in
this paper: 

\begin{definition}[Bimodular form]\label{defjm}
Let $(\Gamma_1,\Gamma_2;\Gamma)$ be a triple of subgroups of $\slr$ commensurable with $\slz$.\footnote{A subgroup $\Gamma\subset \slr$ is commensurable with $\slz$ if $\Gamma\cap \slz$ has finite index in both $\slz$ and $\slr$. This includes in particular all congruence subgroups of $\slz$.}  A two-variable meromorphic function $F:\mathbb
H\times \mathbb H \to
\mathbb C$ is called a bimodular form of weight $(k_1,k_2)$ for the triple
$(\Gamma_1,\Gamma_2;\Gamma)$ if it satisfies both Condition 1 
\& 2:

\begin{itemize}  
\item Condition 1: For all $\gamma_i=\left(\begin{smallmatrix}a_i&&b_i\\c_i&&d_i\end{smallmatrix}\right)\in \Gamma_i$, $i=1,2$,  $F$ transforms as
\begin{equation}
F(\gamma_1\tau_1,\gamma_2\tau_2)=\chi(\gamma_1,\gamma_2)\,(c_1\tau_1+d_1)^{k_1}(c_2\tau_2+d_2)^{k_2}F(\tau_1,\tau_2),
\end{equation}
for a certain multiplier $\chi: \Gamma_1\times\Gamma_2\to \mathbb C^*$. We call this the
\underline{separate} transformation of $F$ under $(\Gamma_1,\Gamma_2)$, and denote it by $(\Gamma_1)_{\tau_1}\times (\Gamma_2)_{\tau_2}$.
\item Condition 2: For all $
\gamma=\left(\begin{smallmatrix}a&b\\c&d\end{smallmatrix}\right)\in\Gamma$, 
 $F$ transforms as
\begin{equation}
F(\gamma\tau_1,\gamma\tau_2)=\phi(\gamma)\,(c\tau_1+d)^{k_1}(c\tau_2+d)^{k_2}F(\tau_1,\tau_2),
\end{equation}
for a multiplier $\phi:\Gamma\to \mathbb C^*$.
We call this the \underline{simultaneous} transformation of $F$ under $\Gamma$, and denote it by $\Gamma_{(\tau_1,\tau_2)}$.
\end{itemize}
\end{definition} 
Note that condition 2 follows from condition 1 if $\Gamma$ is the
intersection of $\Gamma_1$ and $\Gamma_2$, $\Gamma=\Gamma_1\cap
\Gamma_2$ with $\phi(\gamma)=\chi(\gamma,\gamma)$, $\gamma\in\Gamma$.
  
This definition contains the main aspects of other definitions of
bimodular forms in the literature \cite{stienstra_zagier_16,yang2007differential,liuquan2020,Manschot:2021qqe}.

The definition above for the triple
$(\Gamma_1,\Gamma_2;\Gamma_1\cap \Gamma_2)$ is equivalent to
the definition in
\cite{yang2007differential}. For the triple $(\Gamma_1, \Gamma_1;\slz)$, our definition is equivalent to the one of \cite{Manschot:2021qqe}. Finally, for $k_1=k_2$ and the triple $(\Gamma_1, \Gamma_1;\Gamma)$, our definition is equivalent with \cite{liuquan2020}. 
Finally, the definition of Stienstra and Zagier
\cite{stienstra_zagier_16}, as cited in \cite{yang2007differential},
does require Condition 2 without requiring Condition 1.

From definition \ref{defjm}, we find that $\ua: \mathbb H\times\mathbb H\to\mathbb C$ in \eqref{uASW} is a bimodular form of weight $(0,2)$ for the triple 
 \begin{equation}\label{mm00triple}
(\Gamma(2),\Gamma(2);\slz),
\end{equation}
with trivial multipliers $\chi$ and $\phi$. In fact, $m\mapsto  \ua$ is a 1-parameter family of such bimodular forms.

The function \eqref{uASW} can be easily expanded in either $q=e^{2\pi\im\tau}$ or $q_0=e^{2\pi\im\tn}$. When expanding $\ua$ around $q_0=0$, every coefficient is a modular function for $\Gamma(2)_{\tau}$. If we denote the vector space of holomorphic modular forms of weight $k$ for $\Gamma\subseteq\slz$ by $\CM_k(\Gamma)$, then $\ua\in \CM_0(\Gamma(2))\llbracket q_0^{\frac14}\rrbracket$. Conversely, we have that $\ua\in \CM_2(\Gamma(2))\llbracket q^{\frac14}\rrbracket$.

Recall that $\Gamma(2)$ is a genus zero congruence subgroup. As such, its Hauptmodul $\lambda$ is the single transcendental generator of the function field of $\Gamma(2)\backslash\mathbb  H^*$. Since $\ua$ is modular in $\tau$ as well as $\tn$ for $\Gamma(2)$ and no larger subgroup of $\slz$, the transcendence of $\lambda$ then implies that \eqref{uASW} cannot be simplified further. 

The Coulomb branch $\CB_4$ for the mass $\bfm=\bma$ has six singularities that come in three pairs of two. By expanding $\lau$ around the cusps, one easily finds
\begin{equation}\begin{aligned}\label{caseASing}
\ua(\tfrac12,\tn)&=-\frac{m^2}{3}\jt_3(\tn)^4(\luv-2),\\
\ua(0,\tn)&=-\frac{m^2}{3}\jt_3(\tn)^4(\luv+1),\\
\ua(1,\tn)&=-\frac{m^2}{3}\jt_3(\tn)^4(-2\luv+1).
\end{aligned}\end{equation}
Notice that the singularities are holomorphic modular forms of weight
2 for $\Gamma(2)_{\tn}$, and are permuted by elements of $(\slz)/\Gamma(2))_{\tn}$.
The reason for $\ua(\tfrac12,\tn)=\ua(\im\infty,\tn)$ is explained in Section \ref{sec:nf4FD}.
Since $\lambda$ is a Hauptmodul for $\Gamma(2)$, for given $\tn\in \Gamma(2)\backslash\mathbb H$ there is exactly one $\tau\in \Gamma(2)\backslash\mathbb H$ where $u$ has a pole. It is where $\tau$ approaches $\tn$, $\ua(\tn,\tn)=\infty$.

We can furthermore compute the period $\frac{da}{du}$. Actually,
$\frac{da}{du}$ is not invariant under the monodromy around $\infty$, but multiplied by $-1$. Instead of $\frac{da}{du}$, we may consider $\left(\frac{da}{du}\right)^2$, which is monodromy  invariant \cite{Brandhuber:1996ng,aspman2021cutting}. In the pure ($N_f=0$) $\SUT$ case, it is a modular form of weight $2$ for $\Gamma^0(4)$.
The weight is the same in  $N_f=4$, however it also transforms well under fractional linear transformations of $\tn$. More specifically, we find that
\begin{equation}\label{daduA}
\left(\frac{da}{du}\right)_\tta^2(\tau,\tn)=\frac{1}{8m^2}\frac{\jt_3(\tau)^4}{\jt_3(\tn)^8}\frac{\lambda(\tau)-\lambda(\tn)}{\lambda(\tn)(\lambda(\tn)-1)}.
\end{equation}
The normalisation may be checked from the fact that $\frac{da}{du}\sim \frac{1}{\sqrt{8u}}$ for $u\to\infty$, which due to \eqref{uASW} corresponds to $\tau\to\tn$.
Since $\jt_3^4$ is a modular form of weight $2$ for $\Gamma(2)$, it follows that $\left(\frac{da}{du}\right)^2_\tta$ satisfies condition 1 of definition \ref{defjm} with weight $(2,-4)$ for $\Gamma(2)_\tau\times\Gamma(2)_{\tn}$. 
 As $\bfm_\tta$ is left invariant by triality, $\left(\frac{da}{du}\right)^2_\tta$ must also be modular for $\slz$. Indeed, one easily finds that $\left(\frac{da}{du}\right)^2_\tta$ 
also satisfies condition $2$, such that it is a bimodular form of weight $(2,-4)$ for the triple $(\Gamma(2),\Gamma(2);\slz)$.

We can also compute the physical discriminant, which for the case A reads
 \begin{equation}
\Delta_{\text A}=(u-\ua(\tfrac12,\tn))^2(u-\ua(0,\tn))^2(u-\ua(1,\tn))^2. 
\end{equation}
Since the singularities \eqref{caseASing} themselves are modular forms for $\tn$, it is again a bimodular form. One easily computes 
\begin{equation}
\Delta_{\text A}(\tau,\tn)=m^{12}\jt_3(\tn)^{24}\frac{\lambda(\tau)^2(\lambda(\tau)-1)^2\lambda(\tn)^4(\lambda(\tn)-1)^4}{(\lambda(\tau)-\lambda(\tn))^6}.
\end{equation}
As $\jt_3^{24}$ is a modular form of weight $12$ for $\Gamma(2)$, this shows that $\Delta_{\text A}$ has modular weight $(0,12)$ under $\Gamma(2)_\tau\times\Gamma(2)_{\tn}$. With the same reasoning as above, we find that $\Delta_{\text A}$ is a  bimodular form of weight $(0,12)$ for the same triple \eqref{mm00triple}.

\subsection{Case B}\label{sec:caseb}
The equal mass case $\bfm_{\text{B}}=(m,m,m,m)$ can be treated with
the same technique as in the previous subsection. Since $N_f=4$ with
four equal masses flows to $N_f=0$ for $m\to\infty$, we can express
the $\tau$ dependence through the Hauptmodul $\ffc\coloneqq
\frac{\jt_2^4+\jt_3^4}{\jt_2^2\jt_3^2}$ of $\Gamma^0(4)$. In fact, the
$N_f=0$ order parameter \eqref{nf0parameter} is just
$\frac{u}{\Lambda_0^2}=-\frac12\ffc$. The order parameter $\ub$ reads 
\begin{equation}\label{ub}
	\ub(\tau,\tn)=-\frac{m^2}{3}\jt_2(\tn)^2\jt_3(\tn)^2\frac{2\ffc(\tn)^2+\ffc(\tau)\ffc(\tn)-12}{\ffc(\tn)-\ffc(\tau)},
\end{equation}
which thus does not involve $\jt_4$. Since $\jt_2(\tn)^2\jt_3(\tn)^2$ is a holomorphic modular form of weight $2$ for $\Gamma^0(4)_{\tn}$, we find that $\ub(\tau,\tn)$ has bimodular weight $(0,2)$ for the separate transformations under $\Gamma^0(4)_\tau\times \Gamma^0(4)_{\tn}$.

As $\CT:\bmb\mapsto \bmd$, there is no simultaneous action of $T$ on
$\tau$ and $\tn$ leaving $\ub$ invariant. Also, since
$\CS:\bfm_{\text{B}}\mapsto \bfm_{\text{C}}$, $S$ does not leave $\ub$
invariant. However, a subgroup  $\tri_{\bmb}$ of $\mathscr T$ leaves $\bmb$ invariant: Out of the six elements of $\mathscr T$, $\bmb$ is left invariant by $\mathbbm 1$ and $\CT\CS\CT$. As the action of $\mathscr T$ is combined in \eqref{frt} with a simultaneous action on $\tau$ and $\tn$, we find that $\ub$ is expected to be invariant under a simultaneous transformation of $TST\in\slz$. However, due to the algebra \eqref{s3triality}, the same holds for $T^2$. These two matrices generate the congruence subgroup  $\Gamma^0(2)$ of $\slz$. It is straightforward to check from the explicit expression \eqref{ub} that $\ub$ transforms with weight $(0,2)$ under a simultaneous transformation on $\tau$ and $\tn$ of $\Gamma^0(2)_{(\tau,\tn)}$. This proves that $\ub$ is an example of a bimodular form of weight $(0,2)$ for the triple 
\begin{equation}\label{mmmmtriple}
	(\Gamma^0(4),\Gamma^0(4);\Gamma^0(2)).
\end{equation}
As classified in Section \ref{sec:groupaction}, the stabiliser subgroup $\tri_{\bmb}=\{\mathbbm 1, \CT\CS\CT\}$ for the mass $\bmb$ is isomorphic to the group $S_2\cong \mathbb Z_2$  of order 2. This agrees with the fact that $\slz / \Gamma^0(2)\cong  S_2$.

The singularities are 
\begin{equation}\begin{aligned}\label{caseBsing}
		\ub(1,\tn)&=-\frac{m^2}{3}\jt_2(\tn)^2\jt_3(\tn)^2(-\ffc(\tn)), \\
		\ub(0,\tn)&=-\frac{m^2}{3}\jt_2(\tn)^2\jt_3(\tn)^2(2\ffc(\tn)+6),\\
		\ub(2,\tn)&=-\frac{m^2}{3}\jt_2(\tn)^2\jt_3(\tn)^2(2\ffc(\tn)-6),
\end{aligned}\end{equation}
which again are holomorphic modular forms of weight 2 for $\Gamma^0(4)_{\tn}$. Due to the duality group $\Gamma^0(4)_\tau$, we have that $\ub(1,\tn)=\ub(\im\infty,\tn)$. This singularity has degeneracy 4, and flows to $\infty$ for $m\to\infty$. The singularity in the interior is $\ub(\tn,\tn)=\infty$.
One can also check that the singularities \eqref{caseBsing} never collide: The conditions $\ub(1,\tn)=\ub(0,\tn)$ or $\ub(1,\tn)=\ub(2,\tn)$ are equivalent to $f(\tn)=\pm2$, whose only solutions are the two cusps $\tn^+=0$ and  $\tn^-=2$ of $\Gamma(2)$. Since the SW curve is singular for those values of $\tn$, the singularities do not merge for any finite masses.

Similarly as before, one finds
\begin{equation}\label{dadu2B}
	\left(\frac{da}{du}\right)^2_\ttb(\tau,\tn)=\frac{1}{8m^2}\frac{\jt_2(\tau)^2\jt_3(\tau)^2}{\jt_4(\tn)^8} \left(f(\tau)-f(\tn)\right).
\end{equation}
Since $f$ is a Hauptmodul, $\jt_2^2\jt_3^2$  a modular form of weight $2$ and $\jt_4^8$ a modular form of weight 4 for $\Gamma^0(4)$, it follows that $\left(\frac{da}{du}\right)^2_\ttb$ is a bimodular form of weight $(2,-4)$ for the triple \eqref{mmmmtriple}. Finally, the physical discriminant reads
\begin{equation}\label{casebdisc}
	\Delta_{\text B}(\tau,\tn)=m^{12}\jt_4(\tn)^{24}\frac{(f(\tau)^2-4)(f(\tn)^2-4)^2}{(f(\tau)-f(\tn))^6},
\end{equation}
which is a bimodular form of weight $(0,12)$ for the triple \eqref{mmmmtriple}.

\subsection{Case C}\label{sec:casec}
Let us  study the case where only one hypermultiplet is massive,
$\bfm_{\text{C}}=(2m,0,0,0)$.\footnote{The particular normalisation is
  chosen such that the diagram \ref{fig:BCD} holds without any
  prefactors.} Since in the limit $m\to \infty$ we get massless
$N_f=3$, we can express the $\tau$ dependence through the Hauptmodul
$\tilde \ffc=\frac{\jt_3^2\jt_4^2}{(\jt_3^2-\jt_4^2)^2}$ of
$\Gamma_0(4)$. The order parameter of the massless $N_f=3$ theory
reads $\frac{u}{\Lambda_3^2}=-\frac{1}{64}\tilde \ffc$
(\ref{Nf3massless}), and the functions $\ffc$ and $\tilde \ffc$ are
related by $\ffc(4\tau)=16\tilde \ffc(\tau)+2$ (see Appendix
\ref{sec:masslessnf0123}). One finds for the order parameter $\uc$,
\begin{equation}\label{ucnf4}
	\uc(\tau,\tn)=-\frac{m^2}{3}\jt_3(\tn)^2\jt_4(\tn)^2
	\frac{2\tilde \ffc(\tn)^2+(10\tilde \ffc(\tau)+1)\tilde \ffc(\tn)+2\tilde \ffc(\tau)}{\tilde \ffc(\tn)(\tilde \ffc(\tn)-\tilde \ffc(\tau))},
\end{equation}
which is independent of $\jt_2(\tn)$. Again, the factor $\jt_3(\tn)^2\jt_4(\tn)^2$ is a modular form of weight 2 for $\Gamma_0(4)_{\tn}$, and the quotient is a meromorphic modular function of $\Gamma_0(4)$ for both $\tau$ and $\tn$. Thus $\uc$ satisfies condition 1 of definition \ref{defjm} with weight $(0,2)$ for  $\Gamma_0(4)_\tau\times \Gamma_0(4)_{\tn}$. 
 
Since $\CT:\bfm_{\text{C}}\mapsto \bfm_{\text{C}}$, there is a simultaneous $T$-duality
\begin{equation}
	\CT: \uc(\tau+1,\tn+1)=\uc(\tau,\tn),
\end{equation}
which is straightforward to check from \eqref{ucnf4}. As $\CS:\bfm_{\text{C}}\mapsto \bfm_{\text{B}}$, this exchanges the order parameters
\begin{equation}
	\uc(-\tfrac1\tau,-\tfrac{1}{\tn})=\tn^2 \ub(\tau,\tn),
\end{equation}
which we can also explicitly check. We can again study the stabiliser subgroup of $\tri_{\bmc}$ of $\tri$. It is the group generated by $\CT$ and $\CS\CT^2\CS$, such that $\uc$ is expected to transform simultaneously under $T$ and $ST^2S$. These two matrices generate the congruence subgroup $\Gamma_0(2)$ of $\slz$, which is conjugate to $\Gamma^0(2)$. Thus we find that $\uc$ is a bimodular form of weight $(0,2)$ for the triple 
\begin{equation}\label{m000triple}
(\Gamma_0(4), \Gamma_0(4);\Gamma_0(2)).
\end{equation}
Lastly, we can also study
\begin{equation}\label{dadu2C}
	\left(\frac{da}{du}\right)^2_\ttc(\tau,\tn)=\frac{1}{8m^2}\frac{\jt_3(\tau)^2\jt_4(\tau)^2}{\jt_2(\tn)^8}\frac{\tilde f(\tau)-\tilde f(\tn)}{\tilde f(\tau)\tilde f(\tn)}.
\end{equation}
It is straightforward to check that $\left(\frac{da}{du}\right)^2_\ttc$ is a bimodular form of weight $(2,-4)$ for the triple \eqref{m000triple}. For the discriminant $\Delta_{\text C}$ there exists a similar expression to \eqref{casebdisc}, and it is a bimodular form of weight $(0,12)$ for \eqref{m000triple}.

\subsection{Case D}\label{sec:caseD}
Let us finally also study the case $\bfm_{\text{D}}=(m,m,m,-m)$. It is related to cases B and C as in Fig. \ref{fig:BCD}. We have that $\bfm_\ttd\in\CL_{\CS}$, while $\bfm_\ttb\in \CL_{\CS\CT\CS}$ and $\bfm_\ttc\in\CL_{\CT}$. 
From the SW curve one easily finds 
\begin{equation}\label{ud}
\ud(\tau,\tn)=-\frac{m^2}{3}\im \jt_2(\tn)^2\jt_4(\tn)^2\frac{2\hat\ffc(\tn)^2+\hat\ffc(\tau)\hat\ffc(\tn)-12}{\hat\ffc(\tn)-\hat\ffc(\tau)},
\end{equation}
where
\begin{equation}
\hat\ffc(\tau)=f(\tau+1)=\im \frac{\jt_2(\tau)^4-\jt_4(\tau)^4}{\jt_2(\tau)^2\jt_4(\tau)^4}.
\end{equation}
Since $f$ is a Hauptmodul for $\Gamma^0(4)$, $\hat f$ is a Hauptmodul for a subgroup of $\slz$ conjugate to $\Gamma^0(4)$,
\begin{equation}\label{gamma04hat}
\widetilde{\Gamma^0(4)}=T\,\Gamma^0(4)\,T^{-1}=\bra T^4,ST^2\ket.
\end{equation}
A fundamental domain for $\widetilde{\Gamma^0(4)}$ is given by 
\begin{equation}
\widetilde{\Gamma^0(4)}\backslash\mathbb H=\CF\cup T\CF\cup T^2\CF\cup T^3\CF\cup TS\CF\cup T^3S\CF,
\end{equation}
with $\CF=\slz\backslash\mathbb H$.
It is straightforward to check that $\ud(\tau,\tn)$ transforms with weight $(0,2)$ under $\widetilde{\Gamma^0(4)}_\tau\times \widetilde{\Gamma^0(4)}_{\tn}$. 

The subgroup $\tri_{\bmd}\subset\mathscr T$ leaving invariant $\bmd$ is generated by $\CS$ and $\CT^2$. The two corresponding $\slz$ transformations $S$ and $T^2$ generate the theta group $\widetilde{\Gamma^0(2)}\coloneqq \Gamma_\theta$ \eqref{gammatheta}, which is a congruence subgroup of $\slz$ with index 3, conjugate to $\Gamma_0(2)$ and $\Gamma^0(2)$. 
Thus we find that $\ud$ is a bimodular form of weight $(0,2)$ for the triple
\begin{equation}
(\widetilde{\Gamma^0(4)},\widetilde{\Gamma^0(4)};\widetilde{\Gamma^0(2)}).
\end{equation}
The three groups $\{\Gamma_0(2),\Gamma^0(2),\widetilde{\Gamma^0(2)}\}\ni\Gamma$ are in fact the three groups $\slz\supset \Gamma\supset \Gamma(2)$ with index 3 and 2 cusps, and they correspond to the three conjugate order 2 subgroups of $S_3$.

\subsection{Generic mass}\label{sec:genericmass}
The analysis of the A, B, C and D theories may suggest that the order
parameter $u_\bfm$ for a generic mass $\bfm$ transforms with weight
$(0,2)$ under $G_\tau\times G_{\tn}$ for some subgroup $G\subseteq
\slz$. This is however not true in general, as for generic masses
there are branch points and associated branch cuts, which spoil the
modularity \cite{aspman2021cutting}. The discussion in
\cite{aspman2021cutting} for $N_f\leq 3$ suggests that for a fixed
$\tau$ or fixed $\tn$, there is a natural choice of fundamental domain
$\CF(\bfm)\subseteq \mathbb H$ for $u_\bfm$, such that $u_\bfm:
\CF(\bfm)\to \CB_4$ is one-to-one.  For a generic choice of masses, monodromies on
the $u$-plane give rise to monodromies of $\CF(\bfm)$, but these do
not generate a congruence subgroup of $\slz$ for a generic mass. For
special cases however, $\CF(\bfm)$ is equal
to $\Gamma\backslash\mathbb H$ for some subgroup
$\Gamma\subseteq\slz$, such as when $\bfm$ is equal to $\bma$, $\bmb$,
$\bmc$ or $\bmd$, for which  $\Gamma$ is $\Gamma(2)$, $\Gamma^0(4)$,
$\Gamma_0(4)$ or $\widetilde{\Gamma^0(4)}$. If the mass $\bfm$ is such
that $\CB_4$ contains a superconformal Argyres-Douglas point,
$\Gamma\subseteq\slz$ can also be a subgroup of index smaller than $6$
\cite{aspman2021cutting,Closset:2021lhd}. An example of this will be given
in Section \ref{sec:EFG}.

In the above discussed examples A--D, the duality groups $\Gamma_1$ of  $\tau$ and $\Gamma_2$ of $\tn$ are identical. We show in Section \ref{sec:EFG} that this is not generally true, even if $u(\tau,\tn)$ is modular in $\tau$ and $\tn$. However, we can demonstrate that $\Gamma_1\subset \Gamma_2$. A common non-perturbative definition of the UV coupling constant is the low-energy effective coupling $\tau$ in the limit where the order parameter is large, 
\begin{equation}\label{tndefinition}
\tn=\lim_{u\to \infty}\tau(u).
\end{equation}
Since it is not associated with a singularity, it is neither a cusp nor an elliptic point and therefore an arbitrary interior point in the space of $\tau\in\mathbb H$.
If  $\Gamma_1\subsetneq \Gamma_2$ is not a proper subgroup, then in general $\tn\in \Gamma_2\backslash\mathbb H$ is not an element of a choice of  fundamental domain $\Gamma_1\backslash\mathbb H$. However, there exists a $\gamma_1\in\Gamma_1$ with the property that $\gamma_1\tn\in \Gamma_1\backslash\mathbb H$. Since $u(\tau,\tn)$ has weight $0$ in $\tau$, we notice that
\begin{equation}
u(\gamma_1\tn,\tn)=u(\tn,\tn)=\infty,
\end{equation}
which is the weak coupling region in $\CB_4$.
If $\Gamma_2\subsetneq\Gamma_1$ however, then there exist two points $\tn\neq \tilde\tn$ in the fundamental $\Gamma_1\backslash\mathbb H$, which are not related by any element $\gamma_1\in\Gamma_1$. Then $u(\tau,\tn)$ and $u(\tau,\tilde\tn)$ are two distinct points in $\CB_4$. This contradicts the fact that the  $N_f=4$ Coulomb branch $\CB_4$ only contains one such singularity. This shows that indeed $\Gamma_1\subseteq \Gamma_2$.

The weight $(0,2)$ of $u$ can be explained as follows. Monodromies on
the $u$-plane act on the low-energy effective coupling $\tau$ and by
definition leave $u$ invariant. Thus $u(\tau,\tn)$ is required to have
weight $0$ in $\tau$. For $\tn$, recall that the order parameter
relates to the prepotential $F$ of the theory by a logarithmic
derivative with respect to the instanton counting parameter
\cite{Matone:1995rx,Minahan:1997if,Ne,Sonnenschein_1996,Eguchi:1995jh,DHoker:1996yyu} 
\begin{equation}\label{u=dF}
u=4\pi\im q_0\frac{\partial F}{\partial q_0}=2\frac{\partial F}{\partial \tn}.
\end{equation}
As the prepotential $F$ has weight $0$ in $\tn$, this shows that $u(\tau,\tn)$, has weight $2$ in $\tn$.

The other possible modular transformations are those involving the masses, which is the action of the triality group $\spin(8)\rtimes_\varphi \slz$. 
From the above analysis, we  expect that for generic mass $\bfm$ the order parameter $u_\bfm$ transforms as
\begin{equation}\begin{aligned}\label{sttrialityum}
\CT:\quad& u_\bfm(\tau+1,\tn+1)\!\!\!\!&&=u_{\CT\bfm}(\tau,\tn), \\
\CS:\quad&u_\bfm(-\tfrac1\tau,-\tfrac{1}{\tn})&&=\tau_0^2\, u_{\CS\bfm}(\tau,\tn).
\end{aligned}\end{equation}
Due to the branch points and cuts for generic masses, these transformations are again very subtle to perform. 
From  \eqref{s3triality} and in particular $\CT^2=\mathbbm 1$, 
\eqref{sttrialityum} implies 
\begin{equation}
\CT^2:\quad  u_\bfm(\tau+2,\tn+2)=u_{\bfm}(\tau,\tn).
\end{equation}
We can check explicitly that it is true for example for case B as in \eqref{ub}, which is not $\CT$-invariant.

As discussed in Section \ref{sec:groupaction}, the group action $\scrt\times\mathscr M\to\mathscr M$ partitions the mass space $\scrm\ni\bfm$  into three regions $\CL_1$, $\CL_3$ and $\CL_6$, where  the orbits $\scrt \cdot\bfm$ have length 1, 3 and 6. The stabiliser subgroups of $\bfm$ are then subgroups of $S_3$ of order 6, 2 and 1, i.e. isomorphic to  $S_3$, $S_2$ or $S_1=\{e\}$. The homomorphism $\varphi$ \eqref{phihomomorphism}  between $\slz$ and $\tri=\text{Out}(\spin(8))$  then dictates the subgroup 
\begin{equation}
\varphi^{-1}[\tri_{\bfm}]
\end{equation}
under which $u_\bfm$ is simultaneously invariant. The preimage of the stabiliser subgroup under $\varphi$ thus constitutes the third component $\Gamma$ of the triple $(\Gamma_1,\Gamma_2;\Gamma)$ in definition \ref{defjm}.

\subsubsection*{The case $\bfm\in \CL_1$}

When $\bfm\in\CL_1$, then the stabiliser group of $\bfm$ has six elements and the orbit $\scrt \cdot\bfm$ consists of $\bfm$ only. Then there is only one function in \eqref{sttrialityum}, and $u_\bfm$ transforms with weight $(0,2)$ under $\slz_{(\tau,\tn)}$, as in condition 2 of definition \ref{defjm}. An example is  $\ua$ as given in \eqref{uASW}, and the transformation is checked in \eqref{uasl2z}. 

\subsubsection*{The case $\bfm\in \CL_3$}
The case $\bfm\in\CL_3$ is most interesting, as it is not trivial ($\bfm\in\CL_1$) and not generic ($\bfm\in\CL_6$). Namely, when the orbit $\scrt\cdot\bfm$ contains three elements,  the stabiliser group is isomorphic to the symmetric group $S_2$ with two elements. Then the three functions associated with the three elements of the orbit $\scrt\cdot\bfm$ form a vector that transforms under $\slz$. An example for this are the functions $\ub,\uc,\ud$ found in Sections \ref{sec:caseb}--\ref{sec:caseD}. As is clear from  Fig. \ref{fig:BCD}, they are related to each other by triality.  If we organise $\bfu_3=(\ub,\uc,\ud)^\tra$,  using \eqref{ub}, \eqref{ucnf4} \eqref{ud} one can prove that
\begin{equation}\begin{split}\label{t2suaub}
\bfu_3(\tau+1,\tn+1)&=\begin{pmatrix}0&0&1\\0&1&0\\1&0&0\end{pmatrix}\bfu_3(\tau,\tn), \\ \bfu_3(-1/\tau,-1/\tn)&=\tn^2\begin{pmatrix}0&1&0\\1&0&0\\0&0&1\end{pmatrix}\bfu_3(\tau,\tn).
\end{split}\end{equation}
As the matrices are in $\mathrm{GL}(3,\mathbb C)$, there exists a 3-dimensional representation $\slz\to \mathrm{GL}(3,\mathbb C)$. This  shows that $\bfu_3(\tau,\tn)$ furnishes a vector-valued bimodular form of weight $(0,2)$ for $\slz$, agreeing with the following definition:\footnote{It is customary to define vector-valued modular forms for $\slz$, however vector-valued modular forms for proper subgroups $\Gamma$ of $\slz$ are familiar in rational CFTs \cite{Miyamoto1998,krauel2012vertex,Cheng:2020srs} and so we leave our definition more generic.}

\begin{definition}[Vector-valued bimodular form]\label{vvbim}
Let 
\begin{equation}
\bfF=\left(\begin{array}{c}F_1 \\ \vdots \\ F_p\end{array}\right):\mathbb H\times\mathbb H\to \mathbb C^p
\end{equation}
be a $p$-tuple of two-variable meromorphic functions, $p\in\mathbb N$. Then  $\bfF$ is called a vector-valued bimodular form of weight $(k_1,k_2)$ for $\Gamma\subset\slz$, if
\begin{itemize}
\item each component $F_j$ is a bimodular form of weight $(k_1,k_2)$  for some triple $\allowbreak(\Gamma_1^j,\Gamma_2^j; \Gamma^j)$, as in  definition \ref{defjm}, and 

\item there exists a $p$-dimensional complex representation $\rho:\Gamma\to \text{GL}(p,\mathbb C)$ such that
\begin{equation}
\bfF(\gamma\tau_1,\gamma\tau_2)=(c\tau_1+d)^{k_1}(c\tau_2+d)^{k_2}\rho(\gamma)\bfF(\tau_1,\tau_2)
\end{equation}
for all $\gamma=\left(\begin{smallmatrix}a&b\\c&d\end{smallmatrix}\right)\in\Gamma$ and all $\tau_1,\tau_2\in\mathbb H$.
\end{itemize}
\end{definition}

Since $\bfu_3$ is parametrised by the mass $m\in\mathbb C$, $m\mapsto \bfu_3(m,\tau,\tn)$ is in fact a  1-parameter family of vector-valued bimodular forms of weight $(0,2)$ for $\slz$. The triality action of $\slz$ permutes the triples $(\Gamma_1^j,\Gamma_2^j; \Gamma^j)$ in an interesting way. The action of the $\slz$ generators on $u$ is given by \eqref{t2suaub}. As $\Gamma_1^j=\Gamma_2^j$ for the cases B, C, D, both $\Gamma_1^j$ and $\Gamma_2^j$ are conjugated by the corresponding element of $\slz$.  An instance of this is the group $\widetilde{\Gamma^0(4)}$ \eqref{gamma04hat}, which is the set of elements of $\Gamma^0(4)$ conjugated by $T$. Similarly, we have that $\Gamma^0(4)$ is conjugate to $\Gamma_0(4)$ by conjugation with $S$.
The same is true for the three groups $\Gamma^0(2)$, $\Gamma_0(2)$ and $\widetilde{\Gamma^0(2)}$ that the cases B, C, D simultaneously transform under, these three conjugate subgroups are permuted under $\slz$ just as $\Gamma^0(4)$, $\Gamma_0(4)$ and $\widetilde{\Gamma^0(4)}$ are.

\subsubsection*{The case $\bfm\in \CL_6$}
The remaining case is that $\bfm\in \CL_6$, where $\mathscr T\cdot\bfm$ has six elements. Then we can organise $\bfu_6=(u_{\bfm},u_{\CT\bfm},u_{\CS\bfm},u_{\CT\CS\bfm},u_{\CS\CT\bfm},u_{\CT\CS\CT\bfm})^\tra$, which is a collection of six pairwise distinct functions. By studying the action of $\CT$ and $\CS$ on the vector $(\bfm,\CT\bfm,\CS\bfm,\CT\CS\bfm,\CS\CT\bfm,\CT\CS\CT\bfm)^\tra$, we find the transformations
\begin{equation}\begin{aligned}\label{ul6transformation}
\bfu_6(\tau+1,\tn+1)&=&
\begin{pmatrix}
 0 & 1 & 0 & 0 & 0 & 0 \\
 1 & 0 & 0 & 0 & 0 & 0 \\
 0 & 0 & 0 & 1 & 0 & 0 \\
 0 & 0 & 1 & 0 & 0 & 0 \\
 0 & 0 & 0 & 0 & 0 & 1 \\
 0 & 0 & 0 & 0 & 1 & 0 \\
\end{pmatrix}
\bfu_6(\tau,\tn), \\
& & \\
\bfu_6(-1/\tau,-1/\tn)&=\tn^2\!\!\!\!\!&
\begin{pmatrix}
 0 & 0 & 1 & 0 & 0 & 0 \\
 0 & 0 & 0 & 0 & 1 & 0 \\
 1 & 0 & 0 & 0 & 0 & 0 \\
 0 & 0 & 0 & 0 & 0 & 1 \\
 0 & 1 & 0 & 0 & 0 & 0 \\
 0 & 0 & 0 & 1 & 0 & 0 \\ 
\end{pmatrix}
\bfu_6(\tau,\tn). 
\end{aligned}\end{equation}
The vector $\bfu_6$ is not a vector-valued bimodular form for $\slz$,
because the components of $\bfu_6$ do not transform as modular
forms under the separate action of $\Gamma^j_{1,2}$, $j=1,\dots, 6$
due to the branch cuts,
as discussed for the $N_f\leq 3$ theories in
\cite{aspman2021cutting}. 
As we demonstrate in Section \ref{sec:EFG}, the simultaneous action of
$\slz$ on $\tau$ and $\tn$  is not obstructed by the branch cuts of $u(\tau,\tn)$.

The matrices in  \eqref{t2suaub} and \eqref{ul6transformation} are not only in $\mathrm{GL}(n,\mathbb C)$, but they are in fact permutation matrices: Because triality permutes the respective moduli spaces, the order parameters are merely permuted and there are no phases. Thus we have that
\begin{equation}
\bfu(\gamma\tau,\gamma\tn)=(c\tau_0+d)^2P_{\pi(\gamma)}\bfu(\tau,\tn),
\end{equation}
where $P_{\pi(\gamma)}$ is the permutation matrix for the permutation $\pi(\gamma)\in S_{|\mathscr T\cdot\bfm|}$, which can be found from the action of $\mathscr T$ on $\bfm$.

For the period $\left(\frac{da}{du}\right)^2$, there are similar results. For instance, one can check that 
\begin{equation}
\left(\left(\tfrac{da}{du}\right)^2_{\text B},\left(\tfrac{da}{du}\right)^2_{\text C},\left(\tfrac{da}{du}\right)^2_{\text D}\right)^\tra
\end{equation}
is a vector-valued bimodular form of weight $(2,-4)$ for $\slz$. 
As $u$ has weight $(0,2)$, it is not obvious how the discriminant $\Delta$ transforms since it is a polynomial in $u$. However, because triality acts on the $6$ singularities as well, in general $\boldsymbol \Delta$ is a vector-valued bimodular form of weight $(0,12)$ for $\slz$. This can be checked explicitly for the cases B, C, D, where $\boldsymbol \Delta_3=(\Delta_{\text B},\Delta_{\text C},\Delta_{\text D})^\tra$ is a 1-parameter family of vector-valued bimodular forms of weight $(0,12)$ for $\slz$.

\subsection{Cases E, F and G}\label{sec:EFG}
To make the analysis in the previous section more explicit, we can study three cases E, F and G, with $\bfm_{\text{E}}=(m,m,\mu,\mu)$, $\bfm_{\text{F}}=(m+\mu,m-\mu,0,0)$ and $\bfm_{\text{G}}=(m,m,\mu,-\mu)$. These mass vectors share the same symmetry properties, and diagram, as B, C and D (see Fig. \ref{fig:BCD}), i.e. that $\CS$ interchanges E and F, while leaving G invariant and $\CT$ interchanges E and G while leaving F invariant. They also give back all cases A, B, C and D in different limits. For example, if we send $\mu\to 0$ all three cases become case A, while if we send $\mu\to m$ we see that E becomes B, F becomes C and G becomes D. However, due to the fact that we now have two masses, the theories become more complicated. New features such as superconformal fixed points of Argyres-Douglas (AD) type appear, as well as branch points due to square roots \cite{Argyres:1995xn, aspman2021cutting}. 

The order parameters are now given by
\begin{equation}\footnotesize\label{uEFG}
\begin{aligned}
		\ue&=\frac{\jt_3(\tau_0)^4}{6(\lambda-\lambda_0)(\lambda\lambda_0-1)}\Big[ (m^2+\mu^2)(1+\lambda_0)(\lambda_0+\lambda(2+\lambda_0(\lambda-6+2\lambda_0)))\\
		&\quad+3(\lambda^2-1)(\lambda_0-1)\lambda_0\sqrt{(m^2-\mu^2)^2+4m^2\mu^2\frac{\lambda}{\lambda_0}\frac{(\lambda_0-1)^2}{(\lambda-1)^2}}\Big],\\
		\uf&=\frac{\jt_3(\tau_0)^4}{6(\lambda-\lambda_0)(\lambda(\lambda_0-1)-\lambda_0)}\Big[ (m^2+\mu^2)(\lambda_0-2)(\lambda^2(\lambda_0-1)+2\lambda_0^2(\lambda-1))\\
		&\quad+3(\lambda-2)(\lambda_0-1)\lambda_0\sqrt{(m^2-\mu^2)^2\lambda^2+4m^2\mu^2\lambda_0^2\frac{\lambda-1}{\lambda_0-1}}\Big],\\
		\ug&= \frac{\jt_3(\tau_0)^4}{6(\lambda^2-\lambda-\lambda_0^2+\lambda_0)}\Big[ (m^2+\mu^2)(2\lambda_0-1)((\lambda_0-1)\lambda_0+2\lambda^2-2\lambda)\\
		&\quad+3(2\lambda-1)(\lambda_0-1)\lambda_0\sqrt{(m^2-\mu^2)^2+4m^2\mu^2\frac{\lambda}{\lambda_0}\frac{\lambda-1}{\lambda_0-1}}\Big],
\end{aligned}
\end{equation}
where we have abbreviated $\lambda(\tau)=\lambda$, $\lambda(\tau_0)=\lambda_0$. Since $\lambda$ is invariant under $\Gamma(2)$, it naively looks like these order parameters satisfy condition 1 in definition \ref{defjm} of a bimodular form, namely that they transform as bimodular forms under the separate transformations of $\Gamma(2)_\tau\times \Gamma(2)_{\tau_0}$ (or possibly conjugates thereof) with weights $(0,2)$, but due to the presence of the square roots the story is more subtle \cite{aspman2021cutting}. We do, however, find the expected behaviour under the simultaneous action of $\CT$ and $\CS$ for all three cases,
\begin{equation}
	\begin{aligned}
		&\CT:\begin{cases}\ue(\tau+1,\tau_0+1)=&\!\!\!\!\ug(\tau,\tau_0),\\
		\uf(\tau+1,\tau_0+1)=&\!\!\!\!\uf(\tau,\tau_0),\end{cases}\\
		&\CS:\begin{cases}\ue(-\tfrac{1}{\tau},-\tfrac{1}{\tau_0})=&\!\!\!\! \tau_0^2\uf(\tau,\tau_0), \\ \ug(-\tfrac{1}{\tau},-\tfrac{1}{\tau_0})=&\!\!\!\! \tau_0^2\ug(\tau,\tau_0)\end{cases}
	\end{aligned}
\end{equation}
We thus see that, even though the separate action of the duality group
on $\tau$ and $\tn$ becomes more subtle in the presence of square
roots, the simultaneous $\slz$ action involving triality is still
preserved for generic masses. We can also
check the limits to other theories. If we send $\mu\to 0$ we indeed
find that the order parameter of all three cases becomes $\ua$
\eqref{uASW}, consistent with the limit of the mass vector. By sending
$\mu\to m$ we instead find that $\ue\to \ub$, $\uf\to \uc$ and $\ug\to
\ud$ as expected.  

The cusps are given by
\begin{equation}
	\begin{aligned}
		&\text{E}:\begin{cases}u_{m_i}\!\!\!\!&=\frac{\jt_3(\tau_0)^4}{3}(m_i^2(2\lambda_0-1)+m_{j\neq i}^2(2-\lambda_0)), \\ u_\pm\!\!\!\!&=-\frac{\jt_3(\tau_0)^4}{3}((m^2+\mu^2)(\lambda_0+1)\pm 6m\mu\sqrt{\lambda_0}), \end{cases}\\
		&\text{F}:\begin{cases}  u_{m_i}\!\!\!\!&=\frac{\jt_3(\tau_0)^4}{3}(m_i^2(2\lambda_0-1)-m_{j\neq i}^2(\lambda_0+1)),\\ u_\pm\!\!\!\!&=\frac{\jt_3(\tau_0)^4}{3}((m^2+\mu^2)(2-\lambda_0)\pm 6m\mu\sqrt{\lambda_0}), \end{cases}\\
		&\text{G}:\begin{cases}
		u_{m_i}\!\!\!\!&=\frac{\jt_3(\tau_0)^4}{3}(m_i^2(2-\lambda_0)-m_{j\neq i}^2(1+\lambda_0)), \\ u_\pm\!\!\!\!&=\frac{\jt_3(\tau_0)^4}{3}((m^2+\mu^2)(2\lambda_0-1)\pm6m\mu \im \sqrt{\lambda_0(1-\lambda_0)}), \end{cases}
	\end{aligned}
\end{equation}
where $m_i=m,\mu$ and $m_{j\neq i}$ then denotes the other mass. The singularities $u_{m_i}$ all have degeneracy two while $u_\pm$ have degeneracy one. There are also singularities in the interior given by $u(\tau_0,\tau_0)=\infty$ for all three cases.

\subsubsection*{Special points}
As in the theories with $0<N_f\leq 3$ there is a plethora of theories in the moduli space of generic masses $N_f=4$ where the singularity of the fibres is of a higher type, in the sense of Kodaira, and where mutually non-local dyons become massless \cite{Argyres:1995xn}.\footnote{Note however that we should not expect to find any \emph{new} types of theories, compared to the ones of $N_f\leq 3$ in this moduli space, but only types $II$-$IV$ \cite{Argyres:1995xn}. This is because an overall scaling of the masses is not a true parameter of the theory.} These can be classified similarly as in the asymptotically free theories by finding the values of $\bfm$ and $u$ such that $g_2=g_3=0$.  

As we have just seen, there are also theories where the order parameter has branch points due to square roots. A possible interpretation of the corresponding branch points in the asymptotically free theories has been suggested in \cite{aspman2021cutting} as first order phase transitions connected to the second order transition that is the Argyres-Douglas (AD) theories. This implies that we might expect to have branch cuts whenever we have an AD theory. It is straightforward to check that the cases A, B, C and D only have as superconformal fixed points $m\to 0$, $u\to 0$, so that the lack of branch points in these theories is consistent with the above claim. For the more general cases the story changes as we have just seen for cases E, F and G. Let us therefore study the special points of these theories in more detail.

\subsubsection*{AD points}
We define the AD loci as the values of the masses for which there exists an AD theory. This can then be expressed as the zero loci of the polynomials
\begin{equation}\label{AD_EFG}
	\begin{aligned}
		\text{P}^{\text{AD}}_{\text{E}}=&\left(m^2\lambda_0-\mu^2\right)\left(\mu^2\lambda_0-m^2\right)\\
		\text{P}^{\text{AD}}_{\text{F}}=&\left(m^2(\lambda_0-1)+\mu^2\right)\left(\mu^2(\lambda_0-1)+m^2\right), \\
		\text{P}^{\text{AD}}_{\text{G}}=&\left(m^2(\lambda_0-1)-\lambda_0\mu^2\right)\left(\mu^2(\lambda_0-1)-\lambda_0m^2\right),
	\end{aligned}
\end{equation}
Since $T:\lambda\mapsto \frac{\lambda}{\lambda-1}$ and $S:\lambda\mapsto 1-\lambda$ we see that the AD loci also satisfy triality, such that if we act on $\text{P}_{\text{E}}^{\text{AD}}$ with $T$ we get $\text{P}^{\text{AD}}_{\text{G}}$ (up to an overall non-zero factor which is not important since we are looking for the roots of the polynomial) and if we act with $S$ we get $\text{P}^{\text{AD}}_{\text{F}}$.

By tuning the mass to any of the AD values we find that three singularities merge. Depending on which AD mass is chosen, one of the degeneracy two singularities $u_{m_i}$ merge with one of the degeneracy one singularities $u_\pm$. This gives rise to a singular fibre of type $III$ ($\ord(g_2,g_3,\Delta)=(1,2,3)$), implying that three mutually non-local states are becoming massless \cite{Argyres:1995xn}. It is now easy to find closed expressions for $u$ for any of the three theories, and the square roots all disappear.\footnote{Note that for some of the values of the masses the solution we have picked for general $m$ and $\mu$ will become a constant function of $\tau$, this is because the chosen solution corresponds to the solution for $u$ near a singularity that merges with others to become the AD singularity.}

To give an example we take case E and tune the masses such that $m=\mu\sqrt{\lambda_0}$, where $\sqrt\lambda=\frac{\jt_2^2}{\jt_3^2}$ is a holomorphic modular form. The order parameter becomes\
\begin{equation}\label{ueAD}
	\ue^{\text{AD}}=\frac{2\mu^2\jt_3(\tau_0)^4}{3}\frac{(\lambda_0-1)(\lambda_0(\lambda_0(f_2+8(7+\lambda_0)))-56)-8}{\lambda_0(32+f_2-16\lambda_0)-16},
\end{equation}
where $f_2=f_2(\tau)=16\frac{\jt_4(\tau)^8}{\jt_2(\tau)^4\jt_3(\tau)^4}$ is a Hauptmodul of the index 3 congruence subgroup $\Gamma^0(2)\subset \SL$. It is straightforward to check that $\ue^{\text{AD}}$ has weight $0$ under separate transformations for $\Gamma^0(2)_\tau$, and weight $2$ under separate transformations for $\Gamma(2)_{\tn}$. Thus, the group of simultaneous transformations contains $\Gamma^0(2)\cap \Gamma(2)\cong\Gamma(2)$. 
We therefore find that $\ue^{\text{AD}}$ is a bimodular form of weight $(0,2)$ for the triple 
\begin{equation}
(\Gamma^0(2),\Gamma(2);\Gamma(2)).
\end{equation}
Note that this is our first example of a bimodular form that has two different modular groups for the two couplings. The fact that the index in $\SL$ of the modular group of $\tau$ shrinks by the number of merged non-local singularities, $2+1$ in this case, is the expected behaviour of AD theories \cite{Closset:2021lhd,aspman2021cutting}. 

Since the two separate duality groups, $\Gamma^0(2)$ and $\Gamma(2)$,
are different, we cannot choose the fundamental domains for $\tau$ and
$\tau_0$ to coincide as in previous cases. Instead, we can choose the
fundamental domain for $\tau$ as a subset of that for
$\tau_0$. Equation (\ref{ueAD}) demonstrates that $\ue^{\text{AD}}$
has a single pole as a function of $\tau\in \mathbb{H}/\Gamma^0(2)$ for
fixed $\tau_0$,
while it has two poles as function
of $\tau_0\in \mathbb{H}/\Gamma(2)$ for fixed $\tau$. The two points
in $\tau_0\in \mathbb{H}/\Gamma(2)$ are related
by an element in $\Gamma^0(2)/\Gamma(2)$.

We can further note that the AD mass, $\mad=\mu\sqrt{\lambda_0}$, is not invariant under $\Gamma(2)_{\tau_0}$, due to the square root. We rather have that $\mad\to-\mad$ under $T^2$, which is of course another AD point of the theory, and the order parameters of the two theories are given by the same expression. Furthermore, acting with $S$ and $T$ on $\tau_0$ sends this AD mass to the corresponding AD masses of cases F and G, respectively. 

We also have the possibility of tuning $\tau_0$ to a specific value such that more singularities merge. In the above solution, if we fix $\tau_0=1+\im$, or $\lambda_0=-1$, we find that the remaining degeneracy two singularity merge with the degeneracy one singularity such that we get the weight vector $\bfk=(3,3)$. The relation between the masses is now $m=\im \mu$ and the order parameter is actually independent of $\tau$, the curve is simply given by $\CJ=j(\tn)=12^3$. Therefore, the coupling $\tau(u)=\tn=1+\im$ is fixed over the whole Coulomb branch.
This is expected from the same argument as before since we merge two sets of 3 non-local singularities, such that the fundamental domain for $\tau$ just shrinks to a point $\tn$.

\subsubsection*{Branch points}
As previously mentioned, in the more generic cases there will also be branch points. For the theories E, F and G these are given by the branch points of the square roots in \ref{uEFG},
\begin{equation}
	\begin{alignedat}{2}
		\text{E}:& \quad  \frac{\lambda}{(\lambda-1)^2}&&=-\frac{\lambda_0}{(\lambda_0-1)^2}\frac{(m^2-\mu^2)^2}{4m^2\mu^2},\\
		\text{F}:&\quad \frac{\lambda-1}{\lambda^2}&&=\frac{1-\lambda_0}{\lambda_0}\frac{(m^2-\mu^2)^2}{4m^2\mu^2}, \\
		\text{G}:& \quad \lambda(\lambda-1)&&=\lambda_0(1-\lambda_0)\frac{(m^2-\mu^2)^2}{4m^2\mu^2}.
	\end{alignedat}
\end{equation}
In the $u$-plane they are given by
\begin{equation}
	\begin{alignedat}{2}
		\text{E}:& \quad \ubp=&& -\jt_3(\tau_0)^4(1+\lambda_0)\frac{m^4-4m^2\mu^2+\mu^4}{3(m^2+\mu^2)},\\
		\text{F}:&\quad \ubp=&&-\jt_3(\tau_0)^4(\lambda_0-2)\frac{m^4-4m^2\mu^2+\mu^4}{3(m^2+\mu^2)}, \\
		\text{G}:& \quad \ubp=&& \jt_3(\tau_0)^4(2\lambda_0-1)\frac{m^4-4m^2\mu^2+\mu^4}{3(m^2+\mu^2)}.
	\end{alignedat}
\end{equation}
It is straightforward to see that also these points satisfy triality.

\subsection{Fundamental domains}\label{sec:nf4FD}
In the asymptotically free theories we argued that the $u$-planes can be identified with fundamental domains $\CF_{N_f}(\bfm)$ \cite{aspman2021cutting}. For this we make the correspondence that the number of singularities gives the number of rational cusps, the number of BPS states becoming massless at each singularity gives the width of each cusp, and the width at $\im\infty$ is given by $4-N_f$. Then, the sum of all cusps is RG invariant. By following the RG flow from $N_f=3$ to $N_f<3$ we find that  gradually a singularity at strong coupling (a rational cusp) moves to infinity and is identified with the weak coupling region ($\im\infty$). Reversing this argument implies that for $N_f=4$  there should be six rational cusps and the width at infinity vanishes. This is consistent with the fact \cite{Seiberg:1994aj} that $u=\infty$ does not correspond to a cusp of the curve anymore. Rather, it lies in the interior of $\mathbb H\ni\tn$.

\begin{figure}[htbp!]
\centering
	\includegraphics[scale=0.85]{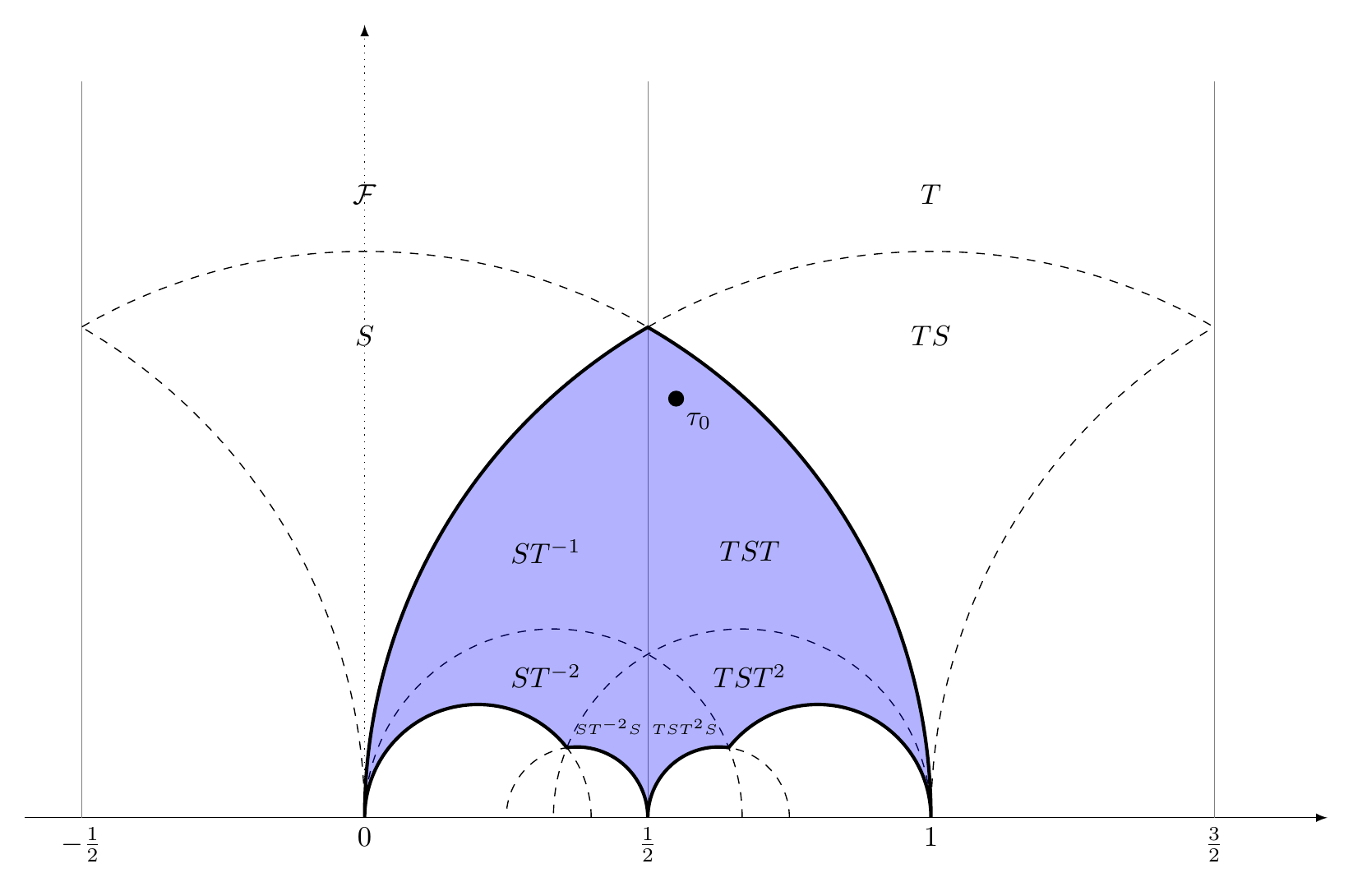}
	\caption{Fundamental domain of $N_f=4$ case A with $\bfm=(m,m,0,0)$.}\label{fig:fund4A}
\bigskip
	\includegraphics[scale=0.85]{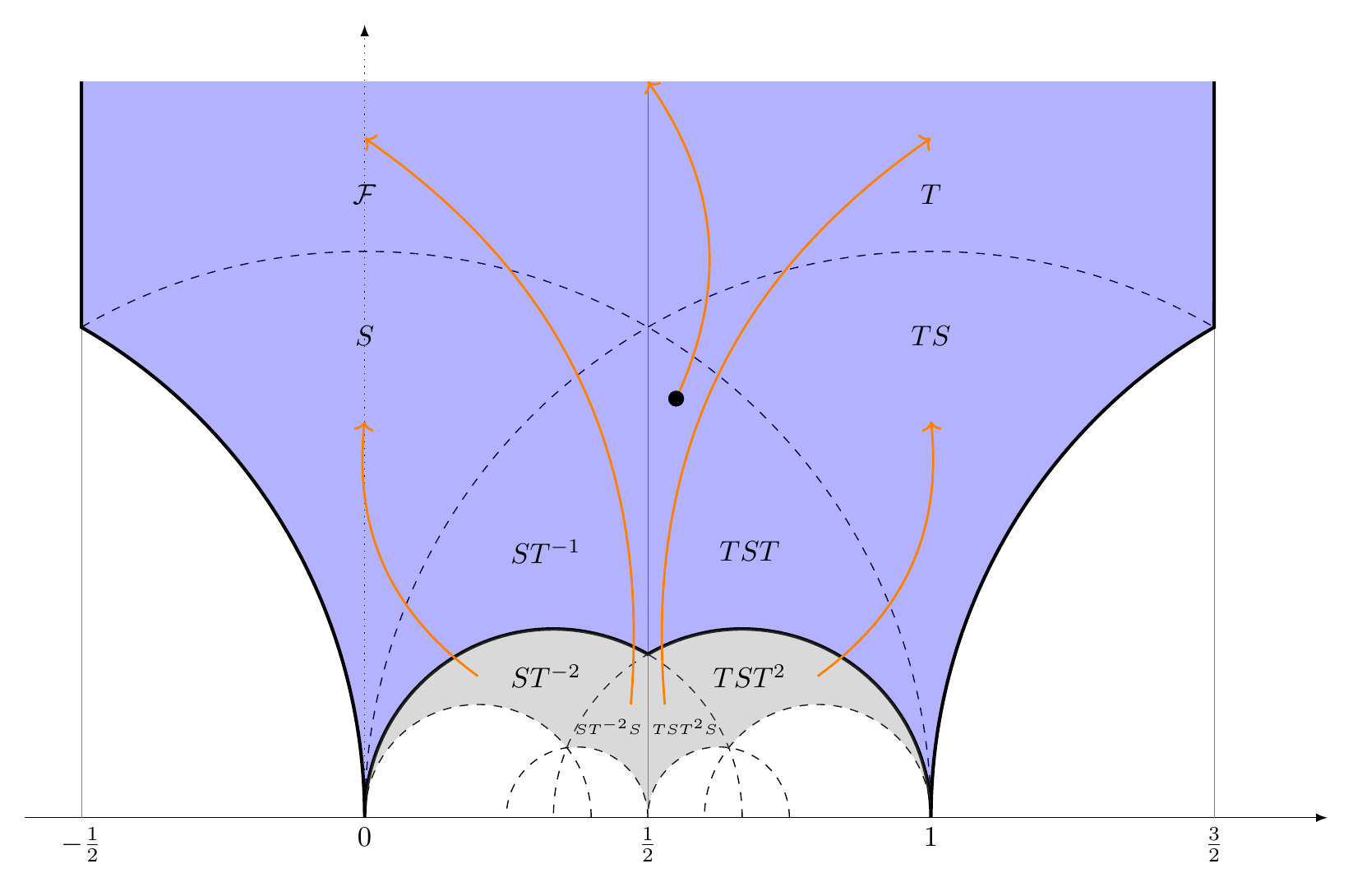}
	\caption{Decoupling the two massive hypermultiplets in $N_f=4$
          case A gives the domain of massless $N_f=2$ (blue). Two of
          the differing regions (gray) are the regions near
          $\tau=\frac12$, which are mapped (orange) to
          $\im\infty$. The remaining two are merely mapped to
          $\Gamma(2)$ equivalent regions near the same cusp such that
          the resulting domain is connected. Alternatively, the fundamental domain
        in this figure is the image of $ST^{-1}STS$ acting on the
        domain in Figure \ref{fig:fund4A}.}\label{fig:fund4A2} 
\end{figure}

It is  found in the above subsections that depending on the mass configuration, the fundamental domain for an order parameter is related to the one of the underlying theory where all massive hypermultiplets are decoupled. We can depict those domains in an equivalent way that is more suitable to our description. For this, one chooses an equivalent fundamental domain with the property that the width at $\im\infty$ is zero and the number of rational cusps is equal to the number of singularities, with according width. For instance, in case A where $\bfm=(m,m,0,0)$ the duality group is $\Gamma(2)$, whose cusps in the decoupling limit (with the same duality group) we choose as $\{\im\infty,0,1\}$. In $N_f=4$ it is more suitable to represent $\im\infty$ by a rational number. For this we can use that $\Gamma(2)\ni ST^{-2}S: \im\infty \mapsto \frac12$, being a preferable  representative of the third cusp. As it necessarily also has width $2$, both $\CF$ and $T\CF$ can be mapped to the region $\tau=\frac12$. This is depicted in Fig. \ref{fig:fund4A}. The decoupling to massless $N_f=2$ is illustrated in Fig. \ref{fig:fund4A2}. The domains in this case are exactly equivalent, Fig. \ref{fig:fund4A} merely allows to extend the $N_f\leq 3$ description of the cusps to $N_f=4$.

We stress that the decoupling limit for $N_f=4$, with the order
parameter a bimodular form as \eqref{uASW}, is quite different from
the asymptotically free theories with $N_f\leq 3$. In the latter
theories, $u(\tau)$ is not holomorphic and modular except for special
points in mass space (complex co-dimension $N_f$)
\cite{aspman2021cutting}. The decoupling of a hypermultiplet is in these
theories accompanied by a branch point moving to infinity. In this way,
a singularity merges with the weak coupling cusp. For cases A, B, C and D in $N_f=4$ on the other hand, there is no branch point for any value of the mass $m$, and in particular there is also none for $m\to\infty$.

\subsection{Relation to $\nstar$}\label{sec:n2star}
Let us collect some results for the  $\mathcal N=2^*$ theory in order to point out the analogy to $N_f=4$. The $\CN=2^*$ theory is obtained by perturbing maximally supersymmetric $\CN=4$ gauge theory by an $\CN=2$ invariant mass term. We have to be careful with the redefinitions. For the $\CN=4$ theory we have $\tau_{\mathrm{0}}=\frac{\theta}{2\pi}+\frac{4\pi\im}{g^2}$ and thus \begin{equation}\label{aaDn=4}
a=\sqrt{2u},\quad a_D=\tau a.
\end{equation}
In this convention we have $u=\frac{a^2}{2}$. This redefinition from the classical formulas  of the theories with fundamental matter is necessary for both representations to feature integral electric-magnetic charges \cite{Seiberg:1994aj}.
The $\nstar$ curve with mass $m$  is identical to the one of $N_f=4$ \eqref{nf4genericcurve} with $\frac12\,\bfm_{\text{A}}=(\frac m2,\frac m2,0,0)$ \cite{Seiberg:1994aj}. A reason for this is that both theories have three singularities each with monodromy being conjugate to $T^2$. This was in fact the ansatz of Seiberg and Witten to determine the curve with generic masses. 

This allows to recycle many results from Section
\ref{sec:caseASWcurve}. The order parameter for $\nstar$ is equal
to\footnote{The expressions for $u_{\nstar}$ in the literature
  \cite{Ferrari:1997gu,laba1998,Huang:2011qx,Bonelli:2019boe,Manschot:2021qqe}
  are related to $\ua(\tau,\tn)$ by a transformation in
  \eqref{sl2zmodgamma2}, which corresponds to the choice of a 
  different solution of the sextic equation associated with the
  $\nstar$ theory \cite{aspman2021cutting}. The different choices can
  be absorbed in the double scaling limit. The counting of the number
  of poles of $\ua$ is immediate from our expression \eqref{uASW} (see comment on the transcendence of $\lambda$ in Section \ref{sec:caseASWcurve}).}
\begin{equation}
u_{\nstar}(\tau,\tn)=\frac14\ua(\tau,\tn).
\end{equation}
 In particular, it is a bimodular form of weight $(0,2)$ for $(\Gamma(2),\Gamma(2); \slz)$.
  The derivative $\frac{da}{du}$ only receives an overall
  normalisation from $N_f=4$, due to \eqref{aaDn=4}. In $\nstar$ the
  singularities each have degeneracy 1 and not 2 as in $N_f=4$ case
  A. Therefore, we have that  $\Delta_{\nstar}=\sqrt{\Delta_{\text{A}}}$,
  which is a polynomial of degree 3 in $u$ and a bimodular form of
  weight $(0,6)$ \cite{Manschot:2021qqe}.

\section{Conclusion and discussion}
In this paper, we have studied in detail the Coulomb branch of the
superconformal $N_f=4$ theory with gauge group SU(2), which has
remained of great interest throughout  the years \cite{Seiberg:1994aj,
  Tai:2010im, Minahan:1996ws, Ferrari:1997gu,
  Dorey:1996bn,Argyres:1999ty,Grimm:2007tm,Huang:2011qx,Tachikawa13,Marino:1998ru,Argyres:1995xn,
  Nekrasov:2015wsu,Jeong:2019fgx,Manschot:2019pog,Nekrasov:2012xe,Gaiotto:2009we,Malmendier:2008db,Gorodentsev:1996cu,Moore:1997pc,Gaiotto:2009hg}. For
the mass configurations with the largest flavour symmetry group, such as when one, two and four
hypermultiplets have an equal mass, we show that the  Coulomb branch
is parametrised by a function $u(\tau,\tn)$ that is not only invariant
under separate modular transformations of $\tau$ and $\tn$, but also
exhibits invariance under a simultaneous transformation under $\tau$
and $\tn$. By restricting to the stabiliser subgroup of a given mass
under the triality action, such order parameters constitute nontrivial
examples of bimodular forms (see \eqref{uASW} for example). Furthermore, the moduli
spaces are permuted under triality, and the order parameter, periods,
discriminants etc. furnish vector-valued bimodular forms, which we
also introduce (see definition \ref{vvbim}). 

The analysis of other mass configurations can be done using the
techniques established in \cite{aspman2021cutting}. As more
complicated mass configurations $\bfm$ inevitably introduce branch
points and cuts, in general $u_\bfm$ is not a bimodular form. A
simultaneous transformation of $\tau$ and $\tn$ is yet to be expected
by triality, while the separate transformations are induced by
monodromies and as such do not in general lie in $\slz$
\cite{Seiberg:1994aj}. However, even in such cases the action of the
monodromy group of the $u$-plane can be understood as paths in the
fundamental domain for $\tau$. See Reference \cite{Aspman:2021vz} for a
discussion of these aspects for gauge group SU(3). 

Our results  allow to study the topologically twisted theory on a four-manifold $X$ \cite{Witten:1988ze,Witten:1995gf,Moore:1997pc, LoNeSha,Laba05}, where the the path integral can be expressed as an integral over the fundamental domain for the effective coupling $\tau$.  In fact, a closed expression for the order parameter is enough to define the integrand. The modularity for $\tau$ allows to show that the integral measure is well-defined. The triality action then gives the S-duality orbit of the $N_f=4$ theory  on $X$ \cite{AFMM:future}.

It would also be interesting to apply our results to other theories
with an IR moduli space of vacua as well as a non-trivial conformal manifold. 
Such theories may include subsectors with triality symmetry, such as  F-theory  \cite{Sen:1996vd}, quiver gauge theories \cite{Gaiotto:2009we}, the AGT correspondence \cite{Alday:2009aq}, little string theory \cite{Duff:1995sm} and string/string/string triality \cite{Bastian:2017ary}.

\acknowledgments
We are happy to thank Cyril Closset, Gregory Moore and Xinyu Zhang for discussions. JA is
supported by the Government of Ireland Postgraduate Scholarship
Programme GOIPG/2020/910 of the Irish Research Council. EF is supported by the TCD
Provost's PhD Project Award. JM is
supported by the Laureate Award 15175 “Modularity in Quantum Field 
Theory and Gravity” of the Irish Research Council.

\appendix

\section{Modular forms}
\label{app:modularforms}
In this appendix, we collect some properties of modular forms for subgroups of $\slz$. For further reading, see \cite{Bruinier08,ono2004,Zagier92,Diamond,schultz2015,koblitz1993}.

We make use of modular forms for the congruence subgroups $\Gamma_0(n)$ and $\Gamma^0(n)$  of $ \slz$. These subgroups are defined as 
\be\begin{aligned}
\Gamma_0(n) = \left\{\begin{pmatrix}a&b\\c&d\end{pmatrix}\in \slz\big| \, c\equiv0 \; \mod n\right\},\\
\Gamma^0(n) = \left\{\begin{pmatrix}a&b\\c&d\end{pmatrix}\in \slz\big| \, b\equiv0 \; \mod n\right\},
\end{aligned}\ee
and are related by conjugation with the matrix $\text{diag}(n,1)$. We furthermore define the \emph{principal congruence subgroup} $\Gamma(n)$ as the subgroup of $\slz\ni A$ with $A\equiv\mathbbm 1\mod n$. A subgroup $\Gamma$ of $\slz$ is called a congruence subgroup if it contains $\Gamma(n)$ for some $n\in \mathbb N$. The smallest such $n$ is then called the \emph{level} of $\Gamma$.

We furthermore make use of the theta group \cite{Bruggeman1994} 
\begin{equation}\label{gammatheta}
\widetilde{\Gamma^0(2)}=\Gamma_\theta\coloneqq\langle T^2,S\rangle\subseteq\slz.
\end{equation}
A fundamental domain for $\Gamma_\theta$ is 
\begin{equation}
\Gamma_\theta\backslash\mathbb H=\CF\cup T\CF\cup TS\CF,
\end{equation}
with $\CF=\slz\backslash\mathbb H$. This
demonstrates that $\Gamma_\theta$ has index $3$ in $\slz$. It is a congruence subgroup of $\slz$, as \cite{conrad,rankin_1977}\footnote{It can also be written as the group of matrices $\left(\begin{smallmatrix}a&b\\c&d\end{smallmatrix}\right)$ with $a+b+c+d\equiv 0\mod  2$, or $ab\equiv cd\equiv 0\mod 2$.}
\begin{equation}
\Gamma_\theta=\left\{A\in\slz\,|\, A\equiv \mathbbm1 \, \text{ or }\, S\mod 2\right\}. 
\end{equation}

The above introduced congruence subgroups host a number of interesting modular forms. 
The Jacobi theta functions $\vartheta_j:\mathbb{H}\to \mathbb{C}$,
$j=2,3,4$, are defined as
\be
\label{Jacobitheta}
\begin{split}
\vartheta_2(\tau)= \sum_{r\in
  \mathbb{Z}+\frac12}q^{r^2/2},\quad 
\vartheta_3(\tau)= \sum_{n\in
  \mathbb{Z}}q^{n^2/2},\quad
\vartheta_4(\tau)= \sum_{n\in 
  \mathbb{Z}} (-1)^nq^{n^2/2},
\end{split}
\ee
with $q=e^{2\pi i\tau}$. These functions transform under $T,S\in \slz$ as
\be
\begin{split}
S:\quad& \begin{array}{l}
  \vartheta_2(-1/\tau)=\sqrt{-i\tau}\vartheta_4(\tau), \\
  \vartheta_3(-1/\tau)=\sqrt{-i\tau}\vartheta_3(\tau), \\ \vartheta_4(-1/\tau)=\sqrt{-i\tau}\vartheta_2(\tau),\end{array}\\
T:\quad& \begin{array}{l}\vartheta_2(\tau+1)=e^{\frac{\pi
      i}{4}}\vartheta_2(\tau), \\
  \vartheta_3(\tau+1)=\vartheta_4(\tau), \\
  \vartheta_4(\tau+1)=\vartheta_3(\tau). \end{array} \label{jttransformations}
\end{split}
\ee
They furthermore satisfy the Jacobi abstruse identity
\begin{equation}\label{jacobiabstruseidentity}
\vartheta_2^4+ \vartheta_4^4= \vartheta_3^4.
\end{equation}
The modular lambda function $\lambda=\frac{\jt_2^4}{\jt_3^4}$ is a Hauptmodul for $\Gamma(2)$.  
The Dedekind eta function $\eta: \mathbb H\to \mathbb C$ is defined as the infinite product
\begin{equation}\label{etatransformation}
\eta(\tau)=q^{\frac{1}{24}}\prod_{n=1}^{\infty}(1-q^n), \quad q=e^{2\pi i\tau}.
\end{equation}
It transforms under the generators of $\slz$ as
\be\begin{aligned}\label{etatransformation}
S: \quad& \eta(-1/\tau)=\sqrt{-i\tau }\, \eta(\tau),\\
T: \quad& \eta(\tau+1)=e^{\frac{\pi i}{12}}\, \eta(\tau),
\end{aligned}\ee
and relates to the Jacobi theta series as $
\eta^{3}=\frac{1}{2}\jt_2\jt_3\jt_4$.

\subsubsection*{Eisenstein series}\label{sec:eisenstein}
We let $\tau\in \mathbb{H}$ and define $q=e^{2\pi i \tau}$. Then the Eisenstein series $E_k:\mathbb{H}\to \mathbb{C}$ for even $k\geq 2$ are defined as the $q$-series 
\be
\label{Ek}
E_{k}(\tau)=1-\frac{2k}{B_k}\sum_{n=1}^\infty \sigma_{k-1}(n)\,q^n,
\ee
with $\sigma_k(n)=\sum_{d|n} d^k$ the divisor sum. For $k\geq 4$ even, $E_{k}$ is a modular form  of weight $k$ for
$\operatorname{SL}(2,\mathbb{Z})$.  Any modular form for $\slz$ can be related to the Jacobi theta functions \eqref{Jacobitheta} by 
\begin{equation}\label{e4e6jacobi}
E_4=\frac12(\jt_2^8+\jt_3^8+\jt_4^8),\qquad E_6=\frac12(\jt_2^4+\jt_3^4)(\jt_3^4+\jt_4^4)(\jt_4^4-\jt_2^4).
\end{equation}
With our normalisation \eqref{Ek} the $j$-invariant can be written as 
\begin{equation}\label{je4e6}
j=1728\frac{E_4^3}{E_4^3-E_6^2}=256\frac{(\jt_3^8-\jt_3^4\jt_4^4+\jt_4^8)^3}{\jt_2^8\jt_3^8\jt_4^8}.
\end{equation}

\section{Partitions of six}\label{app:partitions}
In Section \ref{sec:curve} we listed a number of special mass
configurations of the $N_f=4$ theory, where multiple singularities
have merged. Since the $\CS$ and $\CT$ transformations
preserve the weight vector of the singularities $\bfk(\bfm)$, we could
use this vector as a partial classification of the loci in mass space with merged
singularities. As there are six singularities on the Coulomb branch
$\CB_4$, we can consider partitions of six, of which there are
$p(6)=11$. To further simplify the problem, we consider masses which are either different (but generic), equal or zero. This then means that we should also consider partitions of four for the mass vector. The results are collected in Table
\ref{tab:partsOf6}.

\begin{table}[ht]\begin{center}
\renewcommand{\arraystretch}{1}
		\begin{tabu}{| l|p{4.1cm}|p{3cm}|p{3.5cm}|} 
			\hline 
			$\bfk(\bfm)$ & Example & Singularity Type &
                        Corresponding \newline Section \\ & & & \\ [-1em]
			\hline 
			$(6)$ &$(0,0,0,0)$ & $2I_0^*$  & \text{ \ref{SWcurve}} \\
			$(5,1)$ & - & excluded & - \\
			$(4,2)$ & $(m,m,m,0)$  with \newline
                        $\tau_0=-\tfrac{1}{2}+\tfrac{\im}{2}$ & $I_0^*\,\rom{4}\,II$  & - \\
 			$(4,1,1)$ & $(m,0,0,0)$ & $I_0^*\,I_4\,2I_1$ &
                        B \ref{sec:caseb} , C \ref{sec:casec},
                         D \ref{sec:caseD} \\
 			$(3,3)$ & $(m,m,\mu,\mu)$, \,$\tau_0=1+\im$, 
                        $m=\mu\sqrt{\lambda(\tn)}$, &
                        $I_0^*\,2\rom{3}$ & \ref{sec:EFG} \\
 			$(3,2,1)$ & $(m,m,\mu,\mu)$, \newline 
                         $m=\mu\sqrt{\lambda(\tn)}$ & $\rom{1}_0^*\,\rom{3}\,\rom{1}_2\,\rom{1}_{1}$ &\ref{sec:EFG}  \\
			$(3,1,1,1)$& $(m,m,m,\mu)$, $(m,m,m,0)$ & $\rom{1}_0^*\,\rom{1}_3\,3\rom{1}_1$ & -\\
 			$(2,2,2)$ & $(m,m,0,0)$ & $\rom{1}_0^*\,3I_2$ & \text{A } \ref{sec:caseASWcurve}\\
			$(2,2,1,1)$ & $(m,\mu,0,0)$, $(m,m,\mu,\mu)$ & $\rom{1}_0^*\,2\rom{1}_2\,2I_1$
                        & \text{E, F, G }\ref{sec:EFG}  \\
 			$(2,1,1,1,1)$ & $(m,m,\mu_1,\mu_2)$,
                        $(m,m,\mu,0)$ & $\rom{1}_0^*\,\rom{1}_2\,4\rom{1}_1$
                         & -\\
 			$(1,1,1,1,1,1)$ & $(m_1,m_2,m_3,m_4)$,
                        $(m_1,m_2,m_3,0)$ & $\rom{1}_0^*\,6\rom{1}_1$
                        & -\\ [-1em] &&& \\
			\hline 
		\end{tabu}
		\caption{A partial classification of mass loci of the
                  $N_f=4$ theory, based on the  weight vector
                  $\bfk(\bfm)$. The second column gives one example
                  of the mass vector, where for some values of
                  $\bfk(\bfm)$ it is also necessary to fix $\tn$. Since the weight vector is triality
                  invariant, the orbits $\tri\cdot\bfm$ of $\bfm$
                  under $\mathscr T$ \eqref{scrT} give additional configurations. The
                  third column gives the singularity type following
                  \cite{Persson:1990} for the
                  example in the second column. The
                  last column lists the examples studied in the main text for a given
                  $\bfk(\bfm)$, with a
                  reference to the specific Section. 
}\label{tab:partsOf6}\end{center} 
\end{table}

More generally, we could consider specific relations among
the masses, including $\tau_0$-dependence, such as in the AD theories of
Sec. \ref{sec:EFG}. For these cases, we tuned the mass such that a number of non-local singularities merged and we got a weight vector $\bfk(\bfm_{EFG}^{\text{AD}})=(3,2,1)$. Similar situations appear in the other theories with several mass parameters. For example, in the $(m,m,m,\mu)$ theory with $m=\mad=\frac{\lambda_0}{2-\lambda_0}\mu$ we find that one of the degeneracy one singularities merges with the degeneracy three singularity to give a theory with weight vector $\bfk=(4,1,1)$. We further have the possibility to tune $\tau_0$ to a specific value such that even more singularities merge. For example, in the AD theory of case E studied in Sec. \ref{sec:EFG} we saw that we could fix $\tau_0=1+\im$ to get a theory with weight vector $\bfk=(3,3)$. Similarly, in the $(m,m,m,\mu)$ theory mentioned above, with $m=\mad$ we can set $\tau_0=(-1)^{1/3}$ to find a theory with $\bfk=(4,2)$. In the $(m,m,m,0)$ theory there is no value for $m$ such that more singularities merge (except $m=0$) but we can set $\tau_0=-\frac{1}{2}+\frac{\im}{2}$ to find a $\bfk=(4,2)$ theory. This is the only theory with one mass parameter where this is possible. 
In for example case A, the value of $\tau_0$ needed to merge further singularities is a singular one, corresponding to $\lambda_0=1$, which is not allowed.

As discussed recently in Refs \cite{Caorsi:2018ahl, Closset:2021lhd}, the possible
fibration structures of the SW curve over the $u$-plane for fixed
masses and $\tau_0$ follow the classification of
rational elliptic surfaces by Miranda and Persson \cite{Miranda:1986,
  Persson:1990, Miranda:1990}. Using this classification, each part of
the partition $\bfk(\bfm)$ can be labelled with the appropriate
singularity type. The singularity for $u\to \infty$
($\tau\to \tau_0$) of the $N_f=4$ theory corresponds to the $I_0^*$ singularity of Kodaira's
classification of singular fibres. The cuspidal singularities
correspond to $I_n$, with $n$ the width of the cusp, while the AD
singularities are of type $II$, $III$ and $IV$. Note that a part of the partition can be realized by
different singularity types depending on the charges of the
corresponding massless particles. For example, a part ``2'' can
correspond to either $I_2$ or $II$. Indeed, there are 19 different
configurations with $I_0^*$ \cite{Closset:2021lhd}, while there are only 11 different
partitions of 6. We have included in
the third column in Table \ref{tab:partsOf6} the singularity type for the data in the second
column following Persson's classification \cite[The list]{Persson:1990}.\footnote{We thank Cyril Closset for
pointing out that $\bfk(\bfm)=(5,1)$ is excluded based on the 
classification by Miranda and Persson.}

\section{The  qq-curve}\label{sec:qqcurve}
The parameters $ u$ and $ m_i$ in \eqref{nf4genericcurve} are not immediately related to $\langle \text{tr}\,\phi^2\rangle$ and the masses of the hypermultiplets \cite{Dorey:1996qt,Dorey:1996zj,Dorey:1996bn,Argyres:1999ty,Dorey_1996,Dorey_1996II,Manschot:2019pog}. In fact, there is discrepancy between $u$ and $\langle \text{tr}\,\phi^2\rangle$ for $N_f\geq3$ \cite{Dorey:1996bn}. Let us instead consider the curve obtained from the qq-characters of the $N_f\leq 4$ $\SUT$ theory \cite{Nekrasov:2015wsu,Jeong:2019fgx,Manschot:2019pog,Nekrasov:2012xe}, which is better suited from the perspective of the instanton calculus \cite{NekOk,losev1998}. The elliptic curve for $N_f\leq 4$ in the flat space limit is 
\begin{equation}
	(1-\tq)^2y^{2}=T(x)^2-4\mathtt{q}\prod_{i=1}^{N_{f}}\left(x+m_{i}\right),
\end{equation}
where 
\begin{equation}
	T(x)=\left(1+\mathtt{q}\right)x^{2}+\mathtt{q}x\sum_{i=1}^{N_f}m_{i}-\left(1-\mathtt{q}\right)\ttu.
\end{equation}
In the context of the qq-curve, it is customary to define 
\begin{equation}\label{tqdef}
\tq\coloneqq \lambda(\tn),
\end{equation}
where $\tn$ is the UV coupling.
Expanding out all terms for $N_f=4$ and substituting $(1-\tq)y$ by $y$, we find the quartic curve
\begin{equation}\begin{aligned}\label{qqcurve}
		y^2&=(1-\tq)^2x^4-2 \left\llbracket m_{1}\right\rrbracket (1-\tq)\tq x^3+\left((\left\llbracket m_{1}\right\rrbracket^2+2\ttu)\tq^2-4\left\llbracket m_{1}m_{2}\right\rrbracket \tq-2\ttu\right)x^2\\
		&-2\tq\left(\left\llbracket m_{1}\right\rrbracket (1-\tq)\ttu+2\left\llbracket m_{1}m_{2}m_{3}\right\rrbracket\right)x+(1-\tq)^2\ttu^2-4 \tq\text{Pf}\,\bfm ,
\end{aligned}\end{equation}
with the notation \eqref{bracketnotation} as well as 
\begin{equation}
\left\llbracket m_{1}\right\rrbracket =\sum_{i=1}^{4}m_{i},\quad	\left\llbracket m_{1}m_{2}\right\rrbracket=\sum_{i<j}m_im_j, \quad \left\llbracket m_{1}m_{2}m_3\right\rrbracket=\sum_{i<j<k}m_im_jm_k.
\end{equation}

As opposed to \eqref{nfdecoupling}, the low-energy theory with $N_f<4$ flavours and scale $\Lambda_{N_f}$ is obtained from the qq-curve \eqref{qqcurve} by decoupling $4-N_f$ hypermultiplets with masses $m_j$ in the scaling limit
\begin{equation}\label{dslqq}
	\uv\to\im\infty,\quad m_j\to\infty,\quad  \Lambda_{N_f}^{4-N_f}=4\, \tq \prod_j m_j.
\end{equation}
The order parameter $\ttu$ in the qq-curve \eqref{qqcurve} is not identical to the one of the $N_f=4$ SW curve \eqref{nf4genericcurve} or the one of the $N_f\leq3 $ SW family  \eqref{eq:curves}. The relation between the qq-curve and the $N_f=4$ SW curve is difficult to work out explicitly, however one easily finds that by decoupling hypermultiplets \eqref{dslqq} the order parameters are related through the chain 
\begin{equation}\label{uqq3uSW3}
	\ttu\to u_{N_f=3}-\frac{\Lambda_3}{8}(m_1+m_2+m_3)\to u_{N_f=2}-\frac{\Lambda_2^2}{8}\to u_{N_f=1}\to u_{N_f=0}.
\end{equation}
These constant shifts are likely due to the fact that the instanton partition function is better suited for U(N) gauge theory rather than SU(N).

\subsection{Case 0}\label{case0}

Let us study the curve \eqref{qqcurve} for the case $\bfm=(0,0,0,0)$. Recall the rational functional  $\CR$ \eqref{cjlambda}
which has poles at $p=0,1,\infty$. It relates the Hauptmoduln $j$ of $\slz$ and $\lambda\coloneqq\frac{\jt_2^4}{\jt_3^4}$ of  $\Gamma(2)$ by $\CR(\lambda(\tau))=j(\tau)$. In the massless limit we compute $\CJ(\ttu,\tq,0)=\CR(\tq)$.\footnote{ The invariant of a quartic curve can be found using the formulas given in \cite{Brandhuber:1996ng}, for example. } As we identify $j(\tau)=\CJ$, it follows that the UV-coupling is related to the complex structure of the curve by
\begin{equation}\label{nf4masslessrelation}
	\tq=\lambda(\tau),
\end{equation}
with $\tau$ the low-energy effective coupling and $\tq=\lambda(\tn)$  the UV coupling.
This has already been conjectured in \cite{Grimm:2007tm} by matching calculations in the field theory limit of type IIA string compactification on Enriques Calabi-Yau to the results from the Nekrasov partition function, and was further explored in \cite{Huang:2011qx}. For a more extensive discussion on the non-perturbative finite renormalisation see \cite[Sections 3.4--3.5]{Jaewon:2012wsa}.

In \cite{Gaiotto:2009we}, the parameters of the moduli space of marginal couplings are identified with coordinates on Teichm\"uller spaces of punctured Riemann surfaces. For the case of $\SUT$ theory with $N_f=4$, the corresponding surface is a 4-punctured Riemann sphere, and the natural coordinate is a cross-ratio $\tq$ of the location of the punctures. Under conformal transformations, the punctures are permuted \cite{Huang:2011qx}
\begin{equation}\label{crossratiotransf}
	\tq\sim \left\{\tq,\frac 1\tq,\frac{1}{1-\tq},1-\tq,\frac{\tq}{\tq-1},\frac{\tq-1}{\tq} \right\}.
\end{equation}
We notice that $\tq\mapsto \CR(\tq)$ is invariant under these permutations. The transformations \eqref{crossratiotransf} of the modular lambda function $\lambda$ are generated by $T: \lambda\mapsto \frac{\lambda}{\lambda-1}$ and $S:\lambda\mapsto 1-\lambda$ \cite{chandrasekharan2012elliptic} and form the anharmonic group. Due to the permutations \eqref{crossratiotransf}, we can  view \eqref{nf4masslessrelation} as an identification of the equivalence class
\begin{equation}\label{nf4qmassless}
	\tq\,/\!\sim\,=\lambda(\tau).
\end{equation} 
It is also clear from $\CJ=\CR(\tq)$ that \eqref{nf4masslessrelation} is just one of six solutions \eqref{nf4qmassless}, since the former equation can be written as a polynomial equation of degree 6 in $\tq$ (see also  \cite{aspman2021cutting}). From \eqref{nf4masslessrelation} it is furthermore clear that $\tau=\tn$ is a solution, such that $\tau(u)=\tn$ is constant over the whole Coulomb branch, just as in the case of the SW curve \eqref{0000sol}. 

\subsection{Case A}\label{sec:caseA}
Let us then study the curve for $\bma=(m,m,0,0)$. This theory has global symmetry $\SUT\times \SUT\times \SUT\times \SUO$. The physical discriminant is 
\begin{equation}\label{discnf4m}
	\Delta= (\ttu-\ttu_{0,\tta})^2( \ttu-\ttu_{m,\tta})^2\left( \ttu-\ttu_{*,\tta}\right)^2,
\end{equation}
where  $\ttu_{0,\tta}=0$, $\ttu_{m,\tta}= m^2$, $\ttu_{*,\tta}=\frac{\tq\, m^2}{\tq-1}$.
The Coulomb branch parameter $\ttu$ can be found as described in Section \ref{SWcurve}, 
\begin{equation}\label{uparameternf4m}
	\tua(\tau)=m^2\frac{\tq}{\tq-\lambda(\tau)},
\end{equation}
where again $\lambda$ is a Hauptmodul for $\Gamma(2)$. This shows that $\ttu(\tau,\tq)$ is a modular function for $\Gamma(2)_\tau$. With the definition \eqref{tqdef}, $\tua$ is also a modular function for $\Gamma(2)_{\tn}$. However, a simultaneous $\slz$ transformation on $\tau$ and $\tn$ does not give back the same function. Therefore, $\tua$ satisfies only condition 1 of def. \ref{defjm}, and as such is only invariant under separate transformations  $\Gamma(2)_{\tau}\times \Gamma(2)_{\tn}$.

The singularities \eqref{discnf4m} can be easily associated to the cusps of $\Gamma(2)\backslash\mathbb H$ by expanding the Jacobi theta functions. As opposed to the asymptotically free cases, $\ttu(\cdot,\tq)$ is not weakly holomorphic due to a pole in the interior of $\mathbb H$. In other words, there is a  singularity $\tau=\lambda^{-1}(\tq)$, where $u\to\infty$ at finite coupling. Just as in \eqref{nf4qmassless} and \eqref{tndefinition}, this is essentially a definition of the UV coupling. We can collect, 
\begin{equation}
	\tua(\tfrac12)=\ttu_{m,\tta}, \quad \tua(0)=\ttu_{*,\tta}, \quad \tua(1)=\ttu_{0,\tta}, \quad \tua\left(\lambda^{-1}(\tq)\right)=\infty.
\end{equation}
In the decoupling limit \eqref{dslqq}, we find that $u$ flows to \eqref{unf2m=0} of the massless $N_f=2$ theory, considering the constant shift \eqref{uqq3uSW3}.  

While the order parameter \eqref{uparameternf4m} corresponding to the
qq-curve is significantly simpler than the one  \eqref{uASW} from the
SW curve, the action of triality is obstructed for the qq-curve. Using
the relation \eqref{tqdef}, we can express $\tua$ as a two variable function of
$\tau$ and $\tn$.  Through the single dependence of both $\ua$ and
$\tua$ on $\lambda(\tau)$, we can relate $\tua$ and $\ua$,
\begin{equation}\label{uauqrelation}
\tua(\tau,\tn)=\frac{1}{\jt_4(\tn)^4}\ua(\tau,\tn)-\frac{m^2}{3}\frac{\jt_2(\tn)^4-\jt_4(\tn)^4}{\jt_4(\tn)^4}.
\end{equation}
It is also immediate from this expression that $\tua$ is not
bimodular, as it fails to be invariant under a simultaneous $\slz$
action. However, the obstruction to bimodularity can be expressed analogously to \eqref{uauqrelation}: We have that $\tua(\tau,\tn)=h_1(\tn)\,\ua(\tau,\tn)+h_2(\tn)$ for some meromorphic modular forms $h_j$, while also a simultaneous action of $\slz$ yields
\begin{equation}\begin{aligned}\label{tuasl2z}
\tua(\tau+1,\tn+1)&= \,  \frac{\jt_4(\tn)^4}{\jt_3(\tn)^4}\tua(\tau,\tn)+m^2 \lambda(\tn), \\
\tua(-\tfrac1\tau,-\tfrac{1}{\tn})&=-\frac{\jt_4(\tn)^4}{\jt_2(\tn)^4}\tua(\tau,\tn),
\end{aligned}\end{equation}
or equivalently in terms of $\tq$,
\be
\begin{split}
\tua(\tau+1,\tn+1)&= \,  (1-\tq)\,\tua(\tau,\tn)+m^2\,\tq, \\
\tua(-\tfrac1\tau,-\tfrac{1}{\tn})&=\frac{\tq-1}{\tq}\,\tua(\tau,\tn).
\end{split}
\ee
We thus find that the action of triality is an affine transformation on $\tua$.

\subsection{The other cases}
For the other cases B -- G the story is similar. The order parameters of cases B, C, E and F are listed in Table \ref{tab:nf4BCEFu}. It is straightforward to show that these parameters have the correct flows when decoupling hypermultiplets and the correct limits into each other upon tuning the masses. The singularities are given in Table \ref{tab:nf4BCEFsings} and the corresponding degeneracies can be read off from the physical discriminants, which read
\begin{equation}
	\begin{aligned}
		\Delta_\text{B}=&(\ttu-\ttu_{*,\text{B}})^4(\ttu-\ttu_{+,\text{B}})(\ttu-\ttu_{-,\text{B}}),\\
		\Delta_\text{C} =& \ttu^4(\ttu-\ttu_{+,\text{C}})(\ttu-\ttu_{-,\text{C}}), \\
		\Delta_\text{E}=&(\ttu-\ttu_{m,\text{E}})^2(\ttu-\ttu_{\mu,\text{E}})^2(\ttu-\ttu_{+,\text{E}})(\ttu-\ttu_{-,\text{E}}), \\
		\Delta_\text{F}=&\ttu^2(\ttu-\ttu_{*,\text{F}})^2(\ttu-\ttu_{+,\text{F}})(\ttu-\ttu_{-,\text{F}}).
	\end{aligned}
\end{equation}

\begin{table}[h]\begin{center}
		\centering \setlength{\extrarowheight}{0.4cm}
		\begin{tabular}{|c|c|}\hline
			Theory&$\ttu(\tau)$ \\ \hline
			B& $m^2\left(\frac{1-3\tq}{1-\tq}-\frac{1-\tq}{1-\ffc(\tau)\sqrt\tq+\tq}\right)$  \\
			C& $m^2\frac{\tq^2}{(1-\tq)}\frac{1}{2-\tq-\sqrt{1-\tq}\left(2+1/\tilde\ffc(\tau)\right)}$  \\
			E& $-\frac{2\tq m\mu}{1-\tq}-\frac{\tq(-1+2\tq \lambda-\lambda^2)(m^2+\mu^2)+(1+\lambda)\sqrt \tq\sqrt{\tq (\lambda-1)^2(m^2-\mu^2)^2+4(1-\tq)^2\lambda(m\mu)^2}}{2(\lambda-\tq)(\tq\lambda-1)}$ \\
			F&$ \tq \frac{\frac{\jt_2^8-(\jt_3^8+\jt_4^8)\tq}{1-\tq}\mu^2-(\jt_2^8+2\jt_3^4\jt_4^4\tq)m^2-(\jt_3^4+\jt_4^4)\sqrt{\jt_2^8(m^2-\mu^2)^2+\frac{4\tq^2(m\mu)^2\jt_3^4\jt_4^4}{1-\tq}}}{2(\jt_2^4+\tq \jt_4^4)(\jt_2^4-\tq\jt_3^4)} $ \\ \hline
		\end{tabular}
		\caption{Order parameters for the massive $N_f=4$ cases B, C, E and F.} \label{tab:nf4BCEFu}\end{center}
\end{table}

\begin{table}[h]\begin{center}
		\centering \setlength{\extrarowheight}{0.4cm}
		\begin{tabular}{|c|c|}\hline
			Theory & Singularities \\\hline 
			B & $	\ttu_{*,\text{B}}=m^2\frac{1-3\tq}{1-\tq};\qquad \ttu_{\pm,\text{B}}=2m^2\frac{-2\tq\pm\sqrt{\tq}}{1-\tq}$ \\\hline
			C &$ \ttu_{0,\text{C}}=0;\qquad \ttu_{\pm,\text{C}}= \pm m^2\frac{2(1\pm\sqrt{1-\tq})-\tq}{1-\tq}$ \\\hline
			E& $\ttu_{m,\text{E}}=m^2-\frac{2m\mu\tq}{1-\tq};\qquad \ttu_{\mu,\text{E}}=\mu^2-\frac{2m\mu\tq}{1-\tq};\qquad \ttu_{\pm,\text{E}}=\frac{\tq(m+\mu)^2}{\tq-1}\pm2\sqrt{\tq}\frac{m\mu}{\tq-1}$ \\\hline
			F & $	\ttu_{*,\text{F}}=\tq\frac{\mu^2-m^2}{1-\tq};\qquad \ttu_{\pm,\text{F}}=m^2+\frac{\mu^2}{1-\tq}\mp\frac{2m\mu}{\sqrt{1-\tq}}$ \\\hline 
		\end{tabular}
		\caption{Singularities for the massive $N_f=4$ cases B, C, E and F.} \label{tab:nf4BCEFsings}\end{center}
\end{table}

From Table \ref{tab:nf4BCEFu} it is clear that $\tub(\cdot,\tq)$ is a modular function for $\Gamma^0(4)$ and $\tuc(\cdot,\tq)$ for $\Gamma_0(4)$, in agreement with the results from the SW curve. The singularities of the curves correspond to the cusps of the respective modular curves. The singularities in the interior can be found as 
\begin{equation}\begin{aligned}
		\tub(\tau_{\infty,\ttb})&=\infty, \quad &\tau_{\infty,\ttb}&=\ffc^{-1}\left(\tfrac{1+\tq}{\sqrt\tq}\right),\\
		\tuc(\tau_{\infty,\ttc})&=\infty, \quad &\tau_{\infty,\ttc}&=\tilde\ffc^{\!-1}\left(\tfrac{\sqrt{1-\tq}}{2-2\sqrt{1-\tq}-\tq}\right).
\end{aligned}\end{equation} 
The order parameter of cases E, F and all of the more general cases contain branch points.  For cases E and F they are given by the argument $\tau_{\text{bp}}$ for which the radicands vanish. For case E for instance, it is given by 
\begin{equation}
	f_{2\text{B}}(\tfrac{\tau_{\text{bp,\text{E}}}}{2})=-64\frac{(m\mu)^2}{(m^2-\mu^2)^2}\frac{(1-\tq)^2}{\tq}.
\end{equation}
The AD loci of the theories E, F and G are now given by
\begin{equation}
	\begin{aligned}
		\text{P}^{\text{AD}}_{\text{E}}=&(m^2\tq-\mu^2)(\mu^2\tq-m^2), \\
		\text{P}^{\text{AD}}_{\text{F}}=&(m^2(\tq-1)+\mu^2)(\mu^2(\tq-1)+m^2), \\
		\text{P}^{\text{AD}}_{\text{G}}=&(m^2(1-\tq)+\mu^2\tq)(\mu^2(1-\tq)+ m^2\tq).
	\end{aligned}
\end{equation}
We can note that these polynomials coincide with \eqref{AD_EFG} upon identifying $\tq\coloneqq \lambda_0$. Similarly to \eqref{tuasl2z}, the order parameters $\tub$, $\tuc$ and  $\tud$ do not form a vector-valued bimodular form, which is due to the fact that they transform into each other with shifts in $\tn$.

\section{The asymptotically free theories}\label{sec:masslessnf0123}
The curves for the asymptotically free theories with $0\leq N_f\leq 3$ fundamental hypermultiplets has been determined in \cite{Seiberg:1994aj}. They read
\begin{equation}\label{eq:curves}
\begin{aligned}
N_f=0:\quad y^2=&x^3-ux^2+\frac{1}{4}\Lambda_0^4x, \\
N_f=1:\quad y^2=&x^2(x-u)+\frac{1}{4}m\Lambda_1^3x-\frac{1}{64}\Lambda_1^6, \\
N_f=2:\quad y^2=&(x^2-\frac{1}{64}\Lambda_2^4)(x-u)+\frac{1}{4}m_1m_2\Lambda_2^2x-\frac{1}{64}(m_1^2+m_2^2)\Lambda_2^4, \\
N_f=3:\quad y^2=& x^2(x-u)-\frac{1}{64}\Lambda_3^2(x-u)^2-\frac{1}{64}(m_1^2+m_2^2+m_3^2)\Lambda_3^2(x-u)\\&+\frac{1}{4}m_1m_2m_3\Lambda_3x-\frac{1}{64}(m_1^2m_2^2+m_2^2m_3^2+m_1^2m_3^2)\Lambda_3^2.
\end{aligned}
\end{equation}
By taking the mass of a hypermultiplet to be infinite while sending the dynamical scale of the theory to zero, in the double scaling limit 
\cite{Eguchi1999}
\begin{equation}
m_i\to \infty, \quad \Lambda_{N_f}\to 0, \quad m_i \Lambda_{N_f}^{4-N_f}=\Lambda_{N_f-1}^{4-(N_f-1)}
\end{equation}
the hypermultiplets decouple. Identifying the $\CJ$-invariants of the curves \eqref{eq:curves} with the modular $j$-invariant $j(\tau)$ allows to find the modular $u$-parameters. This was done in \cite{aspman2021cutting} for the massive $N_f=2,3$ curves. As was discussed in \cite{aspman2021cutting} in general there are six different solutions for the order parameter as a function of $\tau$ from the curve. However, there exists a unique choice of solution such that the decoupling limits work out directly. These solutions then also determine the $N_f=4$ solutions uniquely in the same way, by demanding that the decoupling limits work out. 

For $N_f=0$, the unique choice is 
\begin{equation}\label{nf0parameter}
\begin{aligned}
\frac{u}{\Lambda_0^2}=&-\frac{1}{2}\frac{\vartheta_2^4+\vartheta_3^4}{\vartheta_2^2\vartheta_3^2}.
\end{aligned}
\end{equation}
In massless $N_f=1$, the solution is given by
\begin{equation}\label{nf1u}
\begin{aligned}
\frac{u}{\Lambda_1^2}=&-\frac{3}{2^{\frac73}}\frac{E_4^{\frac12}}{(E_4^{\frac32}-E_6)^{\frac13}}.
\end{aligned}
\end{equation}
In equal mass $N_f=2$, one finds
\begin{equation}
\frac{u}{\Lambda_2^2}=-\frac{\jt_4^8+\jt_2^4\jt_3^4+(\jt_2^4+\jt_3^4)\sqrt{16\frac{m^2}{\Lambda_2^2}\jt_2^4\jt_3^4+\jt_4^8}}{8\jt_2^4\jt_3^4},  
\end{equation}
which becomes 
\begin{equation}\label{unf2m=0}
\frac{u}{\Lambda_2^2}=-\frac{1}{8}\frac{\vartheta_3^4+\vartheta_4^4}{\vartheta_2^4}
\end{equation}
in the massless limit. The theory with three hypermultiplets is much more complicated.  For the mass configuration $\bfm=(m,0,0)$ the equations can be solved analogously, and one finds
\begin{equation}
\frac{u}{\Lambda_3^2}=-\frac{2\vartheta_3^4\vartheta_4^4+(\vartheta_3^4+\vartheta_4^4)\sqrt{\frac{64m^2}{\Lambda_3^2}\vartheta_2^8+\vartheta_3^4\vartheta_4^4}}{64\vartheta_2^8}.
\end{equation}
In the massless limit this becomes
\begin{equation}
\label{Nf3massless}
\frac{u}{\Lambda_3^2}=-\frac{1}{64}\frac{\vartheta_3^2\vartheta_4^2}{(\vartheta_3^2-\vartheta_4^2)^2}.
\end{equation}

\providecommand{\href}[2]{#2}\begingroup\raggedright\endgroup

\end{document}